\documentclass[11pt]{article}

\usepackage{bm, commath}
\usepackage{natbib}
\usepackage{chngcntr}
\usepackage{caption}
\usepackage{graphicx}
\usepackage{subfig}
\usepackage{amsmath, amsfonts, amsthm}
\usepackage{float}
\usepackage{longtable}
\usepackage{booktabs,siunitx}
\usepackage{url}
\usepackage{multirow}
\usepackage{amssymb}
\usepackage{blkarray}
\usepackage{bbm}
\usepackage{multicol}
\usepackage{authblk}
\usepackage[shortlabels]{enumitem}
\usepackage{makecell}
\usepackage[section]{placeins}

\usepackage{listings}
\usepackage{bbm}
\usepackage{textcomp}
\usepackage[margin=1in]{geometry}
\usepackage{authblk}
\usepackage[ruled]{algorithm2e}
\SetKwInput{KwParam}{Parameter}
\SetAlgoCaptionLayout{centerline}

\usepackage{sectsty}
\usepackage[doublespacing]{setspace}
\usepackage[utf8]{inputenc}
\usepackage{float}
\usepackage{tikz}
\usetikzlibrary{shapes,decorations,arrows,calc,arrows.meta,fit,positioning}

\newtheorem{theorem}{Theorem}

\newtheorem{assumption}{Assumption}
\newtheorem*{assumption*}{Assumption}

\newtheorem{innercustomgeneric}{\customgenericname}
\providecommand{\customgenericname}{}
\newcommand{\newcustomtheorem}[2]{%
  \newenvironment{#1}[1]
  {%
   \renewcommand\customgenericname{#2}%
   \renewcommand\theinnercustomgeneric{##1}%
   \innercustomgeneric
  }
  {\endinnercustomgeneric}
}

\newcustomtheorem{customthm}{Theorem}
\newcustomtheorem{customassumption}{Assumption}

\usepackage{mathtools}

\usepackage{xcolor}

\newcommand\bz[1]{{\color{black} {#1}}}

\makeatletter
\renewcommand{\algocf@captiontext}[2]{#1\algocf@typo. \AlCapFnt{}#2} 
\def\@algocf@capt@plain{top}
\renewcommand{\algocf@makecaption}[2]{%
  \addtolength{\hsize}{\algomargin}%
  \sbox\@tempboxa{\algocf@captiontext{#1}{#2}}%
  \ifdim\wd\@tempboxa >\hsize
  \hskip .5\algomargin%
  \parbox[t]{\hsize}{\algocf@captiontext{#1}{#2}}
  \else%
  \global\@minipagefalse%
  \hbox to\hsize{\box\@tempboxa}
  \fi%
  \addtolength{\hsize}{-\algomargin}%
}
\makeatother

\begin{document}

\sectionfont{\bfseries\large\sffamily}%

\subsectionfont{\bfseries\sffamily\normalsize}%

\title{\bz{Generalizing the intention-to-treat effect of an active control against placebo} from historical placebo-controlled trials to an active-controlled trial:\\
\Large A case study of the efficacy of daily oral TDF/FTC in the HPTN 084  study}

\author[1]{Qijia He}
\author[2]{Fei Gao}
\author[3]{Oliver Dukes}
\author[4]{Sinead Delany-Moretlwe}
\author[2]{Bo Zhang \thanks{Correspondence to Bo Zhang, Assistant Professor of Biostatistics, Vaccine and Infectious Disease Division, Fred Hutchinson Cancer Center, Seattle, Washington, 98109. Email: bzhang3@fredhutch.org}}

\affil[1]{Department of Statistics, University of Washington}
\affil[2]{Vaccine and Infectious Disease Division, Fred Hutchinson Cancer Center}
\affil[3]{Department of Applied Mathematics, Computer Science and Statistics, Ghent University}
\affil[4]{Wits Reproductive Health and HIV Institute, University of the Witwatersrand, Johannesburg, South Africa.}

\date{}

\maketitle

\noindent

\textsf{{\bf Abstract}: In many clinical settings, an active-controlled trial design (e.g., a non-inferiority or superiority design) is often used to compare an experimental medicine to an active control (e.g., an FDA-approved, standard therapy). One prominent example is a recent phase 3 efficacy trial, HIV Prevention Trials Network Study 084 (HPTN 084), comparing long-acting cabotegravir, a new HIV pre-exposure prophylaxis (PrEP) agent, to the FDA-approved daily oral tenofovir disoproxil fumarate plus emtricitabine (TDF/FTC) in a population of heterosexual women in 7 African countries. One key complication of interpreting study results in an active-controlled trial like HPTN 084 is that the placebo arm is not present and the efficacy of the active control (and hence the experimental drug) compared to the placebo can only be inferred by leveraging other data sources. \bz{In this article, we study statistical inference for the intention-to-treat (ITT) effect of the active control using relevant historical placebo-controlled trials data under the potential outcomes (PO) framework}. We highlight the role of adherence and unmeasured confounding, discuss in detail identification assumptions and two modes of inference (point versus partial identification), propose estimators under identification assumptions permitting point identification, and lay out sensitivity analyses needed to relax identification assumptions. We applied our framework to estimating the intention-to-treat effect of daily oral TDF/FTC versus placebo in HPTN 084 using data from an earlier Phase 3, placebo-controlled trial of daily oral TDF/FTC (Partners PrEP).}%


\section{Introduction}
\subsection{HIV Prevention Trials Network Study 084: A landmark clinical trial in HIV prevention}
\label{subsec: intro HPTN084}
The HIV Prevention Trials Network Study 084 (HPTN 084) is a phase 3, \bz{double-blind}, randomized trial comparing long-acting cabotegravir (CAB-LA), an intramuscular injectable, long-acting form of pre-exposure prophylaxis (PrEP) for HIV prevention, to daily oraltenofovir disoproxil fumarate
plus emtricitabine (TDF/FTC) among HIV-uninfected, heterosexual women \citep{delany2022cabotegravir}. The study was conducted in 7 countries of sub-Saharan Africa, including Botswana, Eswatini, Kenya, Malawi, South Africa, Uganda, and Zimbabwe. Daily oral TDF/FTC (sold under the brand name $\textsf{Truvada}\textsuperscript{TM}$), a World Health Organization (WHO) recommended PrEP for HIV prevention, has been introduced in these countries; however, despite increasing availability and access to oral PrEP in the region, women have faced considerable barriers, including social stigma, judgement and violence \citep{delany2022cabotegravir}, to daily pill-taking, which partly explained why the global HIV prevention efforts have stalled with nearly 1.5 million new HIV infections in 2021, or 4,000 every day, a statistic nearly the same as in 2020. High-risk populations, especially those facing barriers to adhering to the daily oral PrEP, are in urgent need of a long-acting prevention modality like injectible CAB-LA with a dosing schedule of every 8 weeks. HPTN 084 reported an HIV incidence of 0.20 per 100 person-years in the CAB-LA arm compared to 1.86 per 100 person-years in the daily TDF/FTC arm (hazard ratio, 0.12; 95\% CI, 0.05 to 0.31), demonstrating, \emph{unequivocally}, the superiority of CAB-LA compared to the daily oral TDF/FTC (see Figure S5 in Web Appendix E). Not long after this landmark trial, WHO recommended that ``long-acting injectable cabFotegravir (CAB-LA) be offered as an additional HIV prevention option for people at substantial risk of HIV infection" \citep{world2022guidelines}. 

\vspace{-0.2cm}
\subsection{Active-controlled trial; intention-to-treat effect; sources of heterogeneity and bias}\label{subsec: intro NI trial and heterogeneity}
An important aspect of the HPTN 084 study is its adoption of an active-controlled trial design. Active-controlled trials are commonly used in clinical settings to evaluate the safety and effectiveness of an experimental medication compared to a standard therapy (referred to as an active control and abbreviated as AC) when it is \emph{unethical} to randomize patients to placebo and deprive them of the available standard therapies \citep{ellenberg2000placebo}. Two popular choices of active-controlled trial designs are a superiority design and a non-inferiority (NI) design. In an active-controlled trial design (superiority or non-inferiority), the placebo arm is not present, so it is not straightforward to estimate the \emph{intention-to-treat} (ITT) effect of the active control compared to the placebo in the active-controlled trial population \citep{fleming2011some}. 

\bz{There are two motivations for understanding the ITT effect of an active control compared to the placebo in an active-controlled trial. First, in the design stage of a non-inferiority trial, a key design factor is to select a so-called NI margin, defined as an acceptable loss of efficacy comparing the experimental therapy with the AC in the NI trial population. The current standard practice is to set the NI margin to a fraction of the \emph{assumed} ITT effect of the AC; hence, a better understanding of AC's ITT effect facilitates selecting a rigorous and scientifically justifiable NI margin \citep{rothmann2003design,james2003some,fleming2011some}. Second, in a post-hoc analysis of the active-controlled trial data, the ITT effect of AC versus placebo can be used to establish the ITT effect of the experimental drug versus placebo. The ITT effect of an experimental drug plays a key role in designing future trials to evaluate other experimental drugs, where the current experimental drug may serve as active-control comparator.
In addition, it provides evidence to quantify the experimental drug's public health impact and facilitates comparison of the experimental drug to other therapeutics. Lastly, the ITT effect of the experimental drug helps evaluate how much society should be willing to pay for the improved efficacy of the experimental drug compared to AC. For instance, \citet{neilan2022cost} evaluated the cost-effectiveness of CAB-LA using the Cost-Effectiveness of Preventing AIDS Complications model and a key model parameter in this analysis is the ITT effect of CAB-LA versus placebo. What's more, additional HIV prevention modalities, like an HIV vaccine \citep{fauci2017hiv} and monoclonal antibodies \citep{miner2021broadly}, are currently under development. The placebo-controlled intention-to-treat effect of CAB-LA serves as an important benchmark to these new interventions.}

There are at least three sources of heterogeneity that complicate generalizing an AC's ITT effect from any historical, randomized, placebo-controlled trial to the planned active-controlled trial. First, the actual treatment effect of the AC could be heterogeneous (\emph{treatment effect heterogeneity}). \bz{Second, within the same study, different participants could have different probability of adhering to the assigned treatment (\emph{within-trial compliance heterogeneity}); for example, in the field of HIV prevention, it was reported that age was correlated with a person's adherence to the prescribed PrEP dose \citep{grant2014uptake}.} Moreover, the same AC could be implemented differently across trials and even the same participants could respond differently to distinct implementations (\emph{between-trial compliance heterogeneity}). Third, trials could target different populations, and therefore, key demographic and health information could differ among trial populations (\emph{target population heterogeneity}). An interplay among treatment effect heterogeneity, within- and between-trial compliance heterogeneity, and target population heterogeneity may lead to generalization bias \citep{Stuart:2011aa} of the ITT effect. In fact, ITT estimates of the same intervention often differ across historical trials (see, e.g., Table S3 in Web Appendix E). In an editorial discussing discrepancies among these findings, \citet{cohen2012preexposure} concluded:
\begin{quote}
   {\small Why the results differ across the various studies reported to date is unclear. However, important considerations include the populations studied; the likely routes of HIV transmission (vaginal vs. anal mucosa)...and most important, medication adherence by study participants.}
\end{quote}
\citeauthor{cohen2012preexposure}'s \citeyearpar{cohen2012preexposure} comments echo three 
of the aforementioned sources of heterogeneity.

\vspace{-0.2cm}
\subsection{Current FDA guidelines; existing approaches; \bz{related literature;} limitations}
\label{subsec: intro literature review}
Current FDA guidelines for designing an NI trial recommend two strategies for estimating the efficacy of an AC in the planned NI trial from historical evidence \citep{FDA2016}. First, one may choose a historical placebo-controlled trial of the AC and assume that its ITT effect would remain unchanged in the target NI trial. This assumption is known as the ``constancy assumption" \citep{fleming2011some} and is in general implausible considering the various sources of heterogeneity previously discussed. Alternatively, one may employ a meta-analytical approach and derive an \emph{average} estimate based on summary statistics of multiple historical trial results and a random-effects model. The meta-analytical approach acknowledges the variability of ITT estimates across historical trials and incorporates uncertainty quantification using a random effect; however, the method is still largely \emph{ad hoc} and is not underpinned by clear \emph{identification} assumptions. Either way, the FDA guidelines recommend acknowledging the unreliability of generalization and using a ``discounted" estimate as a means of protection against potential generalization bias.

Some authors acknowledge the important role of observed covariates in generalizing the intention-to-treat effect, and have proposed covariate adjustment methods under a ``conditional constancy" assumption, that is, the intention-to-treat effect within the same strata of study participants (defined by their observed covariates) is constant across trials \citep{zhang2009covariate}. \citet{zhang2014sensitivity} develop a sensitivity analysis method that allows for residual inconstancy due to unmeasured confounding after adjusting for observed covariates. The conditional constancy assumption represents a meaningful improvement upon the na\"ive constancy assumption and, to some extent, addresses the target population heterogeneity; however, even the conditional constancy assumption is hard to justify because of the across-trial compliance heterogeneity often arising from different AC implementation strategies. Another unsolved issue concerns unmeasured confounders: What is the precise role of unmeasured confounders in preventing generalization of the ITT effect? 

Under the conditional constancy assumption, recent developments in the generalization and transportation methods for causal inference could be directly leveraged to generalize the ITT from a historical trial to the planned active-controlled trial \citep{Stuart:2011aa,Dahabreh:2019aa}; see, e.g., \citet{degtiar2021review} for a recent review. \bz{\citet[Section 6, Equation 24]{pearl2011transportability} discussed identification of the causal effect in the presence of a post-randomization surrogate endpoint under a sequential ignorability assumption \citep{joffe2009related}. \citet{Kara2016} proposed targeted maximum likelihood estimators (TMLEs) to transport the intention-to-treat effect across populations under a version of the conditional constancy assumption. They also proposed methods to transport the complier average treatment effect across trials. Unlike \citet{Kara2016}, we will attempt to further link the complier average treatment effect back to the intention-to-treat effect of the target population using a novel decomposition of estimand of interest and by leveraging multiple data sources. As a result, the identification functional and the estimators of our target parameter are different from those in \citet{Kara2016}.} More recently, \citet{dahabreh2022generalizing} discuss in detail unidentifiability of the ITT when there are unmeasured common causes of trial participation and treatment, and interpretation of the covariate-standardized ITT estimates (under the conditional constancy assumption) as estimating the effects of joint interventions that scale-up the trial and assign the treatment. \bz{In the absence of patient-level data, many authors have proposed meta-analysis-based approaches to estimating causal effects accounting for noncompliance. For instance, \citet{zhou2019bayesian} proposed a Bayesian hierarchical modeling approach to estimating complier average treatment effects and \citet{zhou2022estimating} proposed a closely related, frequentist approach that targets the same estimand.}

\vspace{-0.2cm}
\subsection{Our contribution}
\label{subsec: intro our contribution}
\bz{In this article, we study how to estimate the ITT effect of an AC in an active-controlled trial using relevant patient-level data from historical placebo-controlled trials of the AC. We adopt the potential outcomes (PO) framework \citep{neyman1923application, rubin1974estimating} and build upon the literature on generalizing causal inference across clinical trials and/or observational studies \citep{Stuart:2011aa, Dahabreh:2019aa} and instrumental variable (IV) methods \citep{AIR1996}.} We discuss different identification assumptions that permit both point and partial identification under this unified framework. One key assumption in our development is a version of the homogeneity assumption extensively discussed in the IV literature (see, e.g., \citealp{swanson2018partial} and references therein), which illuminates the role of unmeasured confounding and disentangles multiple sources of heterogeneity. Our developed framework allows for assessing and quantifying what FDA guidelines refer to as non-statistically-based uncertainties \citep[Page 20]{FDA2016} and places many ``essential considerations" raised in \citet{fleming2011some} in the context of formal causal identification assumptions. Compared to the constancy and conditional constancy assumptions, our identification assumptions are more concrete and could facilitate more informed discussions among stakeholders including scientists, physicians, statisticians, funding agencies and regulators. 


We propose historical-data-driven estimators that point or partially identify the target parameter using relevant historical trials and under different identification assumptions. We assess the finite-sample performance of proposed estimators in simulation studies and discuss strategies for assorted sensitivity analyses. We apply our proposed estimators to estimating the ITT effect of daily oral TDF/FTC against placebo in the HPTN 084 study using data from this trial and an earlier, historical placebo-controlled trial of daily oral TDF/FTC \citep{baeten2012antiretroviral}.

\vspace{-0.2cm}
\section{Notation and framework}\label{sec: notation}
\vspace{-0.2cm}
\subsection{Potential outcomes}
We consider the potential outcomes framework \citep{AIR1996} to formalize a placebo-controlled trial with noncompliance involving an active control (AC) and a placebo (P). Let $Z_i \in \{0, 1\}$ denote a binary treatment assignment ($0$ for placebo and $1$ for AC), and $D_i(Z_i = z_i) \in \{0, 1\}$ the potential treatment received had the unit $i$ been assigned the treatment $Z_i = z_i$. Each study participant has a pre-specified probability of receiving either treatment (AC or P). A study participant with $\{D_i(1), D_i(0)\} = \{1, 0\}$ complies with the treatment assignment and is referred to as a complier. A participant with $\{D_i(1), D_i(0)\} = \{1, 1\}$ is referred to as an always-taker, $\{D_i(1), D_i(0)\} = \{0, 0\}$ a never-taker, and $\{D_i(1), D_i(0)\} = \{0, 1\}$ a defier \citep{AIR1996}. We have assumed the Stable Unit Treatment Value Assumption (SUTVA) in the definition of $D_i(Z_i = z_i)$ so that a study participant's treatment received depends only on the person's own treatment assignment \bz{\citep{rubin1980discussion,AIR1996}}. Each unit is also associated with potential outcomes $\{Y_i(d_i, z_i),~d_i\in \{0, 1\},~z_i\in\{0, 1\}\}$ where we again assume the SUTVA in this definition. Under the exclusion restriction assumption, we further have $Y_i(d_i, z_i) = Y_i(d_i)$, that is, the treatment assignment affects the outcome only via the actual treatment received. Next, we assume $Z$ is randomly assigned and is ``relevant" in the sense that $\mathbb{E}[D(Z = 1) - D(Z = 0)] \neq 0$. The SUTVA, exclusion restriction, relevance, and random assignment will be referred to as ``core IV assumptions" in the rest of the article. \bz{Without imposing additional assumptions, a participant's compliance class, that is, if the person is a complier, an always-taker, a never-taker, or a defier, is not identified from observed data alone. For instance, a participant who was offered daily oral TDF/FTC but did not take it could either be a never-taker or a defier. Some researchers have proposed additional assumptions to further simplify the compliance categories. For instance, \citet{AIR1996} rule out the defiers under the monotonicity assumption that states $D_i(Z_i = 1) \geq D_i(Z_i = 0)$. Another typical assumption is to assume that those assigned placebo cannot cross over to the active treatment so $D_i(Z_i = 0) = 0$; this setting is referred to as ``one-sided noncompliance" in the literature and exclude defiers and always-takers. We do not {\it a priori} make these additional assumptions, though we will consider these as important special cases.}

\bz{We use the indicator $S$ to denote the trial membership of a participant: $S = t$ if a participant is in the target active-controlled trial; $S = h$ if a participant is in a generic historical placebo-controlled trial. In the later development, we will also explore scenarios where data from two historical trials may be leveraged; in this case, we will use $h_1$ and $h_2$ to distinguish distinct historical trials. Regardless of trial membership, each participant is associated with a vector of baseline covariates $\mathbf{X}$. We will use $\mathcal{P}_{\text{t}}$ to denote the joint distribution of $\mathbf{X}$ in the target trial and $\mathcal{P}_{\text{h}}$ that in the historical trial $h$. We use $\mathbb{E}_{\mathbf{X}\in \mathcal{P}_{\text{t}}}[\cdot]$ and $\mathbb{E}_{\mathbf{X}\in \mathcal{P}_{\text{h}}}[\cdot]$ to denote taking expectation over $\mathcal{P}_t$ and $\mathcal{P}_h$.}

\vspace{-0.2cm}
\subsection{Estimands}
\bz{We define estimands of interest using the notation introduced above. The conditional intention-to-treat effect of AC versus P in the target active-controlled trial is defined as $ITT(\mathbf{X}; S = \text{t}) = \mathbb{E}[Y(Z = 1) - Y(Z = 0) \mid \mathbf{X}, S = \text{t}]$. Averaging the stratum-specific effect $ITT(\mathbf{X}; S = \text{t})$ over the distribution of observed covariates $\mathbf{X} \in \mathcal{P}_{\text{t}}$ then yields the average intention-to-treat effect below:
\begin{equation}\label{eqn: ITT}\small
   ITT(S = \text{t}) = \mathbb{E}_{\mathbf{X}\in \mathcal{P}_{\text{t}}}\left[\mathbb{E}\left[Y(Z = 1) - Y(Z = 0) \mid \mathbf{X}, S = \text{t}\right]\right],
\end{equation}
which is of primary scientific interest and hence our \emph{target parameter}. As discussed in Section \ref{subsec: intro NI trial and heterogeneity}, the NI margin and the ITT effect of the experimental drug can be immediately determined once $ITT(S = \text{t})$ is determined. In parallel, we use $ITT(\mathbf{X}; S = \text{t}) = \mathbb{E}[Y(Z = 1) - Y(Z = 0) \mid \mathbf{X}, S = \text{h}]$ to denote the conditional $ITT$ effect of the AC in the historical placebo-controlled trial. The average $ITT$ effect of the AC in the historical trial is then obtained by averaging the stratum-specific $ITT$ effect over the historical trial population as follows:
\begin{equation}\label{eqn: ITT(S = h)}\small
   ITT(S = h) = \mathbb{E}_{\mathbf{X}\in \mathcal{P}_{\text{h}}}\left[\mathbb{E}\left[Y(Z = 1) - Y(Z = 0) \mid \mathbf{X}, S = h\right]\right],
\end{equation}
which was unbiasedly estimated in the historical placebo-controlled trial by virtue of randomization.}

\bz{Two estimands are closely related to the $ITT$ effect estimand. The average treatment effect, also known as the causal `per-protocol effect' \citep{hernan2017per}, is defined as $ATE:= \mathbb{E}[Y(D = 1) - Y(D = 0)]$ and describes the average treatment effect of drug uptake, rather than drug assignment, on the outcome. Unlike $ITT$, $ATE$ is in general not identified even in a randomized, placebo-controlled trial because, unlike $Z$, $D$ is not randomized. In the HPTN 084 study, the $ITT$ effect is more clinically relevant than the ``per-protocol" effect because ``noncompliance" or ``non-adherence" to daily pills is an important feature of daily oral prophylaxis and the key motivation to developing long-acting prevention modalities like CAB-LA, monoclonal antibodies \citep{miner2021broadly} and a HIV vaccine \citep{fauci2017hiv}. Another related estimand is the complier average treatment effect, which describes the ATE among the latent complier subgroup under the monotonicity assumption \citep{AIR1996}. Compared to the ITT or ATE, the complier average treatment effect is in general more challenging to be generalized because the complier group is latent and could be altered under different circumstances.}

\subsection{The constancy assumption}
The constancy assumption in the NI trial literature \citep{fleming2011some} states the following relationship between the ITT effect of AC in a target NI trial and that in a chosen historical trial:
\begin{assumption}[Constancy]
\label{ass: constancy assumption}
Let $ITT(S = \text{t})$ and $ITT(S = h)$ be defined as in \eqref{eqn: ITT} and \eqref{eqn: ITT(S = h)}, respectively. The constancy assumption is said to hold if $ITT(S = \text{t}) = ITT(S = h)$.
\end{assumption}
Another version of the constancy assumption, referred to as the \emph{conditional constancy assumption} \citep{zhang2009covariate}, states the following:
\begin{assumption}[Conditional constancy]
\label{ass: conditional constancy assumption}
Let $\mathbf{X}$ denote a vector of observed covariates collected in the planned NI trial, and $ITT(\mathbf{X}; S = \text{t})$ and $ITT(\mathbf{X}; S = h)$ denote conditional intention-to-treat effects in the planned NI trial and the chosen historical trial, respectively. Then the conditional constancy assumption is said to hold if $ITT(\mathbf{X}; S = \text{t}) = ITT(\mathbf{X}; S = h)$.
\end{assumption}

It is transparent from definitions \eqref{eqn: ITT} and \eqref{eqn: ITT(S = h)} that when trials enroll study participants from different populations, that is, when $\mathcal{P}_{\text{t}} \neq \mathcal{P}_{\text{h}}$, then Assumption \ref{ass: constancy assumption} could fail even when Assumption \ref{ass: conditional constancy assumption} holds. This has been discussed in great detail in the context of \emph{generalizability and transportability} by many authors; see, e.g., \citet{Dahabreh:2019aa}. Under Assumption \ref{ass: conditional constancy assumption}, the intention-to-treat effect in the planned active-controlled trial is identified from the observed data of the historical trial $S = h$, and may be estimated using outcome regression, inverse-participation-weighting, or a doubly-robust combination of both; see, e.g., \citet[Section 5]{Dahabreh:2019aa}. Although weaker than Assumption \ref{ass: constancy assumption}, Assumption \ref{ass: conditional constancy assumption} is still a strong assumption; it is, after all, a statement about the \emph{intention-to-treat} effect, not the actual treatment effect. Even when both interventions (AC and P) are expected to have consistent treatment effects for similar study participants across different trials, interventions may be implemented differently, induce different compliance even among similar study participants, and lead to different \emph{intention-to-treat} effects. This is particularly true in HIV prevention studies with daily oral PrEP \citep{cohen2012preexposure}.

\bz{Finally, an assumption closely related to conditional constancy is described in \citet{Kara2016}: $\mathbb{E}(Y|D,Z,\mathbf{X}, S=\text{t}) = \mathbb{E}(Y|D,Z,\mathbf{X}, S = h)$. Unlike Assumption \ref{ass: conditional constancy assumption}, this also conditions on $D$ (treatment taken). Although this may sometimes be more plausible than standard conditional constancy, it is nevertheless difficult to interpret on its own, given that it concerns the observed variables rather than potential outcomes. Indeed, if there is any unmeasured treatment-outcome confounding, it is not an assumption about the stability of the causal effect of $D$ on $Y$ but rather the non-causal conditional association, which may be harder to motivate.}

\vspace{-0.2cm}
\section{Identification assumptions; a road map for estimation and sensitivity analysis}
\bz{In this section, we replace the constancy assumption with a set of assumptions regarding effect homogeneity and generalizability that would permit either point or partial identification of the target parameter. These assumptions are not necessarily weaker than the constancy assumption; however, they are transparent, problem-specific, and more amenable to being assessed and critiqued. They also motivate the estimation procedures and associated sensitivity analyses.}

\vspace{-0.2cm}
\subsection{No-interaction/homogeneity-type assumption}\label{subsec: no-interaction assumption}
Intuitively, a statement about the intention-to-treat effect implicitly entails a statement about the compliance structure (that is, how people comply or not comply with the prescribed treatment) and the actual treatment effect (that is, the effect of the treatment had the study participant taken the treatment). To establish a more revealing relationship, we employ some version of a homogeneity or no-interaction assumption. There are several no-interaction-type identification assumptions that help link the intention-to-treat effect to the average treatment effect; see, e.g., \citet[Section 5]{swanson2018partial}. Below, we adopt one version from \citet{wang2018bounded}.

\begin{assumption}[No-interaction]\label{ass: no interaction}
Let $U$ denote unmeasured covariates that confound $D$'s effect on $Y$. The no-interaction assumption holds if there is no additive $U$-$D$ interaction in $\mathbb{E}[Y(D) \mid \mathbf{X}, U]$:
\begin{equation}
    \label{eqn: no-interaction 1}\small
    \mathbb{E}[Y(D = 1) \mid \mathbf{X}, U] - \mathbb{E}[Y(D = 0) \mid \mathbf{X}, U] = \mathbb{E}[Y(D = 1) \mid \mathbf{X}] - \mathbb{E}[Y(D = 0) \mid \mathbf{X}]\tag{Assumption 3a}.
\end{equation}
\textbf{or} no additive $U$-$Z$ interaction in $\mathbb{E}[D(Z) \mid \mathbf{X}, U]$: 
\begin{equation}
    \label{eqn: no-interaction 2}\small
    \mathbb{E}[D(Z = 1)\mid \mathbf{X}, U] - \mathbb{E}[D(Z = 0) \mid \mathbf{X}, U] = \mathbb{E}[D(Z = 1) \mid \mathbf{X}] - \mathbb{E}[D(Z = 0) \mid \mathbf{X}]\tag{Assumption 3b}
\end{equation}
\end{assumption}

\bz{Assumption \ref{ass: no interaction} holds if either Assumption 3a or Assumption 3b holds. Assumption 3a holds if there are no more modifiers of $D$'s effect on $Y$ beyond those captured by $\mathbf{X}$. Assumption 3a does \emph{not} hold, for instance, if some genetic factor is suspected to modify $D$'s effect on $Y$.} \citet{fleming2011some} describe an example where the effect of epidermal growth factor receptor-inhibiting drugs in colorectal cancer patients depends strongly on whether tumors express the wild type or the mutated version of the KRAS gene. In this example, the KRAS gene ($U$) modifies the effect of drug ($D$) on colorectal cancer ($Y$), and Assumption 3a \emph{fails} in an analysis not accounting for it. 

\bz{Assumption \ref{ass: no interaction} also holds if Assumption 3b holds, that is, when the unmeasured modifier of $D$'s effect on $Y$ does not interact with the treatment assignment $Z$ in predicting the treatment received.} In the colorectal cancer example, Assumption \ref{ass: no interaction} would still stand if the KRAS gene does \emph{not} interact with a colorectal cancer patient's treatment assignment in predicting whether or not the patient adheres to the prescribed treatment conditional on the observed covariates $\mathbf{X}$ (possibly including some easier-to-measure aspects of the tumor). This appears to be more reasonable, at least in some applications. 


\bz{Assumption 3 is a generic assumption that could be applied to either the target AC trial $S = t$ or a historical trial $S = h$.} Assumption \ref{ass: no interaction}, when applied to the hypothetical placebo-controlled trial in the planned active-controlled trial population, implies the following decomposition:
\begin{equation}
\begin{split}
\label{eqn: decomposition}\small
    &ITT(\mathbf{X}; S = t) = \underbrace{\mathbb{E}[Y(D = 1) - Y(D = 0) \mid \mathbf{X}, S = t]}_{CATE(\mathbf{X}; S = t)} \times \underbrace{\mathbb{E}[D(Z = 1) - D(Z = 0) \mid \mathbf{X}, S = t],}_{CC(\mathbf{X}; S = t)}
    \end{split}
\end{equation}
where the term $CATE(\mathbf{X}; S = t)$ describes the average treatment effect of AC versus P conditional on a study participant's covariates, and the conditional compliance term $CC(\mathbf{X}; S = t)$ describes the effect of treatment assignment on treatment received conditional on a study participant's covariates, both in the planned active-controlled trial. 

\vspace{-0.3cm}
\subsection{Conditional average treatment effect; mean generalizability}\label{subsec: CATE term}
To link the active-controlled trial to historical trials, we
make the following mean generalizability (also known as mean exchangeability) assumption \citep{Stuart:2011aa, Dahabreh:2019aa}:
\begin{assumption}[Mean generalizability/exchangeability]
    \label{ass: mean generalizability}
    $CATE(\mathbf{X}; S = t) := \mathbb{E}[Y(D = 1)-Y(D = 0) \mid \mathbf{X}, S = t] = \mathbb{E}[Y(D = 1) - Y(D = 0) \mid \mathbf{X}, S = h] := CATE(\mathbf{X}; S = h)$.
\end{assumption}
Assumption \ref{ass: mean generalizability} essentially says that study participants with the same covariate profile $\mathbf{X}$ would experience the same average treatment effect of $D$ on $Y$ in the hypothetical trial and the selected historical trial $S = h$. The major difference between Assumption \ref{ass: mean generalizability} and Assumption \ref{ass: conditional constancy assumption} is that Assumption \ref{ass: mean generalizability} is a statement about the \emph{actual treatment effect} rather than the \emph{intention-to-treat effect}. Assumption \ref{ass: mean generalizability} is in some sense the minimal assumption needed to extend inference from a historical trial cohort to a target population \citep[Section 3]{Dahabreh:2019aa}.

Assumption \ref{ass: mean generalizability} can still be violated if there exists multiple versions of an active control or placebo across trials (i.e., Rubin's SUTVA is violated); \bz{for instance, this could happen if the active control therapy employed in a historical trial is different from that in the planned active-controlled trial due to difference in dosage or ancillary therapies \citep{fleming2011some,FDA2016}. For Assumption \ref{ass: mean generalizability} to hold, researchers should select a historical trial whose active control is as similar as possible to the planned active-controlled trial (e.g., both investigating the same medication at the same dose with near-identical ancillary therapies).} It is important to note that Assumption \ref{ass: mean generalizability} only requires that the AC therapy itself is identical between trials, not the methods of implementation or dissemination. A sensitivity analysis that models $CATE(\mathbf{X}; S = t)$ as a fraction of $CATE(\mathbf{X}; S = h)$ should be considered when Assumption \ref{ass: mean generalizability} is suspected not to hold.

Identification of the conditional average treatment effect from a historical placebo-controlled trial, i.e., $CATE(\mathbf{X}; S = h)$, has been discussed extensively in the literature. Two identification strategies are available. First, \emph{point identification} could be achieved by further imposing Assumption \ref{ass: no interaction} on the selected historical trial. Alternatively, $CATE(\mathbf{X}; S = h)$ is \emph{partially identified} under different sets of identification assumptions, including minimal, core IV assumptions (see, e.g., \citet{swanson2018partial} for a recent review). 

\vspace{-0.3cm}
\subsection{Conditional compliance}\label{subsec: conditional compliance term}
The conditional compliance term is a difference between $CC_{AC}(\mathbf{X}; S = t) := \mathbb{E}[D(Z = 1)\mid \mathbf{X}, S = t]$ and $CC_{P}(\mathbf{X}; S = t) := \mathbb{E}[D(Z = 0) \mid \mathbf{X}, S = t]$. It then suffices to identify each term separately. The former term equals $\mathbb{E}[D\mid \mathbf{X}, S = t, Z = 1]$ and is identified based on the compliance data from the active-controlled trial by virtue of randomization. The latter term $CC_{P}(\mathbf{X}; S = t)$ is not identified from the active-controlled trial, but may be estimated using relevant historical trial data under the following placebo-arm compliance generalizability assumption:
\begin{assumption}[Placebo-arm compliance generalizability/exchangeability]
    \label{ass: one-arm compliance generalizability}
    $\mathbb{E}[D(Z = 0) \mid \mathbf{X}, S = t] = \mathbb{E}[D(Z = 0) \mid \mathbf{X}, S = h]$.
\end{assumption}
\noindent Note that $\mathbb{E}[D(Z = 0) \mid \mathbf{X}, S = h] = \mathbb{E}[D \mid \mathbf{X}, S = h, Z = 0]$ is directly estimable from historical trial data by randomization. Alternatively, researchers may do a sensitivity analysis for $CC_{P}(\mathbf{X})$ by varying it from $0$ to a sensible value. \bz{For instance, if the active control therapy is not available in the AC trial population, then one would reasonably set $CC_{P}(\mathbf{X}) = 0$.} Researchers may also vary $CC_{P}(\mathbf{X})$ in a sensitivity interval centered around $\mathbb{E}[D\mid \mathbf{X}, S = h, Z = 0]$. Either way, instead of outputting a point estimate of $CC_{P}(\mathbf{X})$ and $CC(\mathbf{X})$, one may output a plausible range of them.


In applications where the intention-to-treat effect of AC is an important design factor, e.g., when selecting the NI margin in the design phase of the NI trial, researchers do not have compliance data from the NI trial and need to identify the entire conditional compliance term using data from a historical trial under the following mean compliance generalizability assumption:
\begin{assumption}[Mean compliance generalizability/exchangeability]
    \label{ass: mean compliance generalizability}
    $\mathbb{E}[D(Z = 1) - D(Z = 0) \mid \mathbf{X}, S = t] = \mathbb{E}[D(Z = 1) - D(Z = 0) \mid \mathbf{X}, S = h]$.
\end{assumption}
\noindent By randomization of $Z$, the quantity
$\mathbb{E}[D(Z = 1) - D(Z = 0) \mid \mathbf{X}, S = h]$ equals $\mathbb{E}[D\mid \mathbf{X}, S = h, Z = 1] - \mathbb{E}[D\mid \mathbf{X}, S = h, Z = 0]$ and
is directly estimable from historical trial data. In order for Assumption \ref{ass: mean compliance generalizability} to hold (or approximately hold), the selected historical trial should have a near-identical implementation strategy of the AC as in the active-controlled trial. Researchers are also advised to relax Assumption \ref{ass: mean compliance generalizability} in a sensitivity analysis that varies the conditional compliance term in a sensitivity interval around $\mathbb{E}[D(Z = 1) - D(Z = 0) \mid \mathbf{X}, S = h]$.

\vspace{-0.3cm}
\subsection{Summary of identification strategies and non-statistically-based uncertainties}\label{subsec: summary of identification}

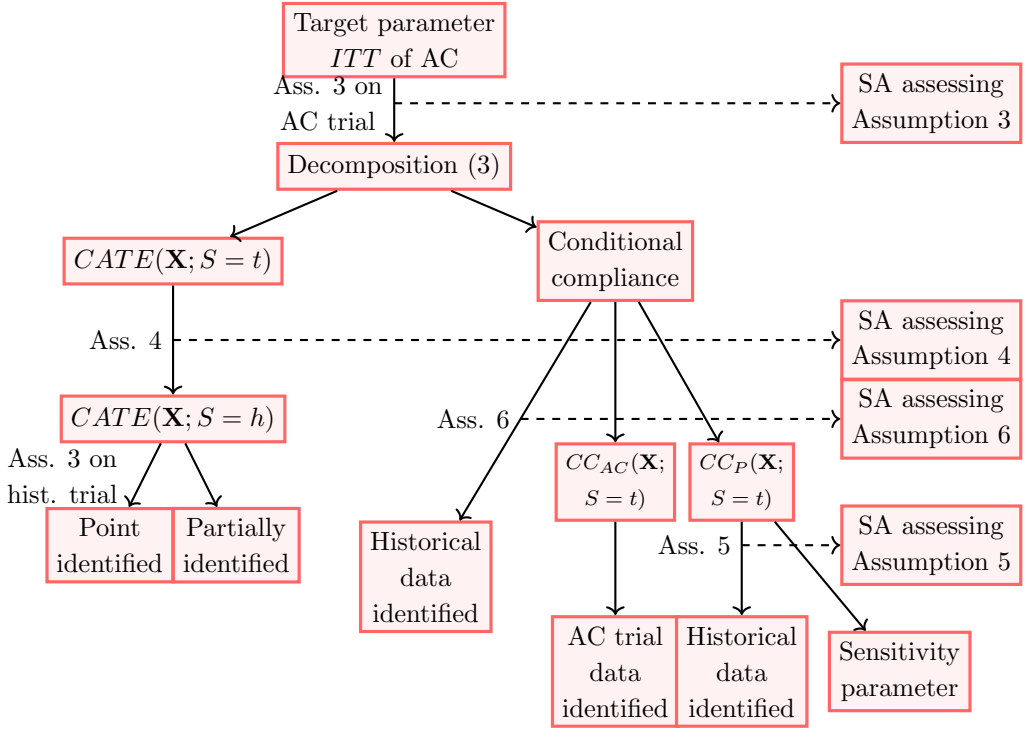
\begin{figure}[ht]
    \centering
    \begin{tikzpicture}[scale = 0.42,
squarednode/.style={rectangle, draw=red!60, fill=red!5, very thick}, 
every text node part/.style={align=center}
]
\node[squarednode] at (1, 4) (target) {\small Target parameter \\ \small $ITT$ of AC};
\node at (-1.1, 2) (Assumption 3) {\small Ass. \ref{ass: no interaction} on\\ \small AC trial};
\node[squarednode] at (1, 0) (decomp) {\small Decomposition \eqref{eqn: decomposition}};
\node[squarednode] at (18, 2) (SA 1) {\small SA assessing \\ \small Assumption \ref{ass: no interaction}};
\node[squarednode] at (-6, -3) (CATE) {\small $CATE(\mathbf{X}; S = t)$};
\node[squarednode] at (8, -3) (CC) {\small Conditional \\ \small compliance};
\node[squarednode] at (-6, -8) (CATE_h) {\small $CATE(\mathbf{X}; S = h)$};
\node at (-7.5, -5.5) (Assumption 4) {\small Ass. \ref{ass: mean generalizability}};
\node[squarednode] at (18, -5.5) (SA 2) {\small SA assessing \\ \small Assumption \ref{ass: mean generalizability}};
\node[squarednode] at (-8, -12) (CATE point) {\small Point \\ \small identified};
\node at (-9.5, -9.8) {\small Ass. \ref{ass: no interaction} on \\ \small hist. trial};
\node[squarednode] at (-4, -12) (CATE partial) {\small Partially \\ \small identified};
\node[squarednode] at (8, -10) (CC term1) {\scriptsize $CC_{AC}(\mathbf{X};$ \\\scriptsize $S = t)$};
\node[squarednode] at (12, -10) (CC term2) {\scriptsize $CC_{P}(\mathbf{X};$ \\\scriptsize $S = t)$};
\node[squarednode] at (8, -16) (CC term1 strategy) {\small AC trial \\\small data \\\small identified};
\node[squarednode] at (12, -16) (CC term2 strategy) {\small Historical \\ \small data \\\small identified};
\node[squarednode] at (17, -16) (CC term2 sens) {\small Sensitivity \\ \small parameter};
\node[squarednode] at (2, -13) (CC strategy) {\small Historical \\ \small data \\\small identified};
\node at (3.5, -8) (Assumption 4) {\small Ass. \ref{ass: mean compliance generalizability}};
\node[squarednode] at (18, -8) (SA 3) {\small SA assessing \\ \small Assumption \ref{ass: mean compliance generalizability}};
\node at (10.5, -12) (Assumption 5) {\small Ass. \ref{ass: one-arm compliance generalizability}};
\node[squarednode] at (18, -12) (SA 4) {\small SA assessing \\ \small Assumption \ref{ass: one-arm compliance generalizability}};
\draw[thick, ->] (target) -- (decomp);
\draw[thick, dashed, ->] (1, 2) -- (SA 1);
\draw[thick, ->] (decomp) -- (CATE);
\draw[thick, ->] (decomp) -- (CC);
\draw[thick, ->] (CATE) -- (CATE_h);
\draw[thick, dashed, ->] (-6, -5.5) -- (SA 2);
\draw[thick, ->] (CATE_h) -- (CATE point);
\draw[thick, ->] (CATE_h) -- (CATE partial);
\draw[thick, ->] (CC) -- (CC term1);
\draw[thick, ->] (CC) -- (CC term2);
\draw[thick, ->] (CC) -- (CC strategy);
\draw[thick, ->] (CC term1) -- (CC term1 strategy);
\draw[thick, ->] (CC term2) -- (CC term2 strategy);
\draw[thick, ->] (CC term2) -- (CC term2 sens);
\draw[thick, dashed, ->] (5, -8) -- (SA 3);
\draw[thick, dashed, ->] (12, -12) -- (SA 4);

\end{tikzpicture}
    \caption{\small A schematic flow chart summarizing different identification assumptions, quantities involved in the estimation, mode of identification, and associated sensitivity analyses examining core assumptions.}
    \label{fig: scheme flowchat}
\end{figure}

Figure \ref{fig: scheme flowchat} in Web Appendix summarizes four aspects we have discussed so far: (i) identification assumptions, including those necessary to identify causal quantities in a trial with non-compliance, and those necessary to generalize inference across trials, (ii) quantities involved in the estimation procedure, (iii) different modes of identification, including point versus partial identification, and identification using historical trials data alone versus using historical trials data and partial active-controlled trial data, and (iv) sensitivity analyses relaxing core assumptions. Together, they help quantify what FDA guidelines refer to as ``\emph{non-statistically-based uncertainties}" \citep[Page 20]{FDA2016}. We next discuss \emph{statistically-based uncertainties}, that is, those associated with sampling variability, by formally proposing estimators for the target parameter.

\vspace{-0.3cm}
\section{Estimation and inference}
\bz{We consider two scenarios for estimation and inference, each corresponding to one major scientific objective of estimating the ITT effect of the AC versus placebo. We first consider the design stage where researchers have access to data from the historical trial $S = h_1$, $\mathcal{D}_{h_1} = \{(\mathbf{X}_i, Z_i, D_i, Y_i, S_i = h_1): i=1,\dots, N_1\}$, data from a second historical trial $S = h_2$, $\mathcal{D}_{h_2} = \{(\mathbf{X}_i, Z_i, D_i, S_i = h_2): i= N_1 + 1,\dots, N_1 + N_2\}$, and baseline covariates data from the target AC trial $\mathcal{D}_t = \{(\mathbf{X}_i, S_i = t) : i = N_1 + N_2 + 1,\dots, N_1 + N_2 + N\}$. Researchers will attempt to leverage the historical data in $S = h_1$ to estimate $CATE(\mathbf{X}; S = t)$ and data in $S = h_2$ to estimate $CC(\mathbf{X}; S = t)$. In this scenario, we write $\mathcal{D} = \mathcal{D}_{h_1} \cup \mathcal{D}_{h_2} \cup \mathcal{D}_t$, where $|\mathcal{D}|$ denote its cardinality. We next consider a more stylistic case where the interest lies in estimating the ITT effect of the AC and hence the experimental therapy versus placebo in an \emph{post hoc} analysis after seeing the compliance data from the target AC trial. In this case, researchers would have access to data $\mathcal{D}_{h_1}$ as mentioned previously plus data $\mathcal{D}_t = \{(\mathbf{X}_i, Z_i, D_i, S_i = t) : i = N_1 + 1,\dots, N_1 + N\}$ from the target AC trial. In this second scenario, we write $\mathcal{D} = \mathcal{D}_{h_1} \cup \mathcal{D}_t$.}

\subsection{\bz{Estimation and inference in the design stage}}
\label{subsec: estimation in the design stage}
We first consider the task of estimating the causal estimand $ITT(S = t)$ in the design stage and derive a simple, regression-based, historical-data-driven estimator under Assumption \ref{ass: no interaction} (for both $S = t$ and $S = h_1$), Assumption \ref{ass: mean generalizability} and Assumption \ref{ass: mean compliance generalizability}. Under Assumption \ref{ass: no interaction}, the conditional average treatment effect in the trial $S = h_1$, i.e., $CATE(\mathbf{X}; h_1)$, is identified as follows:
\begin{equation}
\label{eqn: ratio estimator of CATE in S = h}
\begin{split}\small
    \mathbb{E}[Y(D = 1) - Y(D = 0) \mid \mathbf{X}, S = h_1] &= \frac{\mathbb{E}[Y\mid \mathbf{X}, S = h_1, Z = 1] - \mathbb{E}[Y\mid \mathbf{X}, S = h_1, Z = 0]}{\mathbb{E}[D\mid \mathbf{X}, S = h_1, Z = 1] - \mathbb{E}[D\mid \mathbf{X}, S = h_1, Z = 0]} \\
    &= \frac{\delta^Y(\mathbf{X} ; h_1)}{\delta^D(\mathbf{X} ; h_1)},
\end{split}
\end{equation}
where 
$$
\begin{aligned}\small
\delta^Y(\mathbf{X} ; s) &=\mathbb{E}[Y \mid \mathbf{X}, S=s, Z=1]-\mathbb{E}[Y \mid \mathbf{X}, S=s, Z=0], \\
\delta^D(\mathbf{X} ; s) &=\mathbb{E}[D \mid \mathbf{X}, S=s, Z=1]-\mathbb{E}[D \mid \mathbf{X}, S=s, Z=0].\\
\end{aligned}
$$
\noindent \bz{The expression \eqref{eqn: ratio estimator of CATE in S = h} is sometimes known as the \textit{conditional Wald estimand} \citep{wang2018bounded}.} It also identifies the conditional complier average treatment effect, if we further assume \emph{monotonicity} in the historical trial \citep{AIR1996}. Suppose that we obtain $\hat{\delta}^Y(\mathbf{X}; h_1)$ and $\hat{\delta}^D(\mathbf{X}; h_1)$ by fitting correctly specified parametric models for $\mathbb{E}[Y\mid \mathbf{X},S=h_1, Z=z]$ and $\mathbb{E}[D\mid \mathbf{X},S=h_1, Z=z]$ and that these models are indexed by finite-dimensional parameters which are estimated, for instance, via standard maximum likelihood. Then a regression-based estimator $\widehat{CATE}(\mathbf{X}; h_1)$ is obtained as $ \widehat{CATE}(\mathbf{X}; h_1) = \hat{\delta}^Y(\mathbf{X}; h_1)/{\hat{\delta}^D(\mathbf{X}; h_1)}$. As discussed by \citet{wang2018bounded}, a limitation of this approach with a binary outcome is that one may obtain estimates of $CATE(\mathbf{X};h_1)$ outside of the $[-1,1]$ interval. A regression-based estimator of conditional compliance in the historical trial $h_2$ can be analogously obtained as $\widehat{CC}(\mathbf{X}; h_2) = \hat{\delta}^D(\mathbf{X};h_2)$,
where $\hat{\delta}^D(\mathbf{X};h_2)$ denotes an estimator for the unknown $\delta^D(\mathbf{X};h_2)$ obtained from $\mathcal{D}_{h_2}$ via parametric regression modelling of $\mathbb{E}[D\mid \mathbf{X}, S=h_2, Z=z]$. By averaging $\widehat{CATE}(\mathbf{X}; h_1)$ and $\widehat{CC}(\mathbf{X}; h_2)$ over $\mathbf{X} \in \mathcal{D}_{t}$, we obtain the following regression-based estimator of $ITT(S = t)$:
\begin{equation}\label{eqn: reg-based ITT estimator}
\begin{split}\small
    \widehat{ITT}_{\text{full, reg}} &= \frac{1}{|\mathcal{D}_{t}|} \sum_{i = 1}^{|\mathcal{D}|} \mathbbm{1}\{S_i = t\} \times \left\{\widehat{CATE}(\mathbf{X}_i; h_1) \times \widehat{CC}(\mathbf{X}_i; h_2)\right\} \\
    &= \frac{1}{|\mathcal{D}_{t}|} \sum_{i = 1}^{|\mathcal{D}|} \mathbbm{1}\{S_i = t\}\times \frac{\hat{\delta}^Y(\mathbf{X}_i;h_1)}{\hat{\delta}^D(\mathbf{X}_i;h_1)}\hat{\delta}^D(\mathbf{X}_i;h_2).
\end{split}
\end{equation}
By standard M-estimation theory  \citep{stefanski2002calculus}, the estimator $\widehat{ITT}_{\text{full, reg}}$ is a consistent and asymptotically normal estimator for the target parameter $ITT(S = t)$ under identification and modeling assumptions previously discussed. To obtain a confidence interval, one may use an empirical sandwich variance estimator or the non-parametric bootstrap \citep{cheng2010bootstrap}.

The regression-based estimator $\widehat{ITT}_{\text{full, reg}}$ is expected to perform well if parametric models are correctly specified. Below, we describe an estimator derived from semiparametric efficiency theory \citep{bickel1993efficient}, which allows for more flexible estimation of nuisance functions using modern statistical learning approaches, whilst still facilitating parametric-rate inference on the target parameter. It is developed from the same general theory as recent developments in \textit{de-biased machine learning} \citep{chernozhukov2017double} and \textit{targeted learning} \citep{van2011targeted}.

Recall that the target parameter $ITT(S = t)$ can be expressed as the functional 
$$\small
\psi=\mathbb{E}\left[\frac{\delta^Y(\mathbf{X} ; h_1)}{\delta^D(\mathbf{X} ; h_1)}\delta^D\left(\mathbf{X} ; h_2\right) \mid S = t\right].
$$

\noindent Theorem \ref{speff} gives our main result on semiparametric inference.

\begin{theorem}\label{speff}
Under a non-parametric model $\mathcal M$ that places no restrictions on the observed data distribution, the \textit{efficient influence function} (EIF) for $\psi$ is equal to 
$${\small
\begin{aligned}
EIF_{\psi}=&\frac{1}{\kappa} \frac{(2 Z-1)  \mathbbm{1}\left(S=h_1\right)}{f\left(Z \mid \textbf{X}, S=h_1\right)} \frac{f(S = t \mid \mathbf{X})}{f\left(S=h_1 \mid \mathbf{X}\right)} \frac{\delta^D\left(\mathbf{X} ; h_2\right)}{\delta^D\left(\mathbf{X} ; h_1\right)}\left[Y-\mu_{Y,0}\left(\mathbf{X} ; h_1\right)-\left\{D-\mu_{D,0}\left(\mathbf{X} ; h_1\right)\right\} \frac{\delta^Y(\mathbf{X} ; h_1)}{\delta^D(\mathbf{X} ; h_1)}\right] \\
&+\frac{1}{\kappa} \frac{(2 Z-1)  \mathbbm{1}\left(S=h_2\right)}{f\left(Z \mid \textbf{X}, S=h_2\right)} \frac{f(S = t \mid \mathbf{X})}{f\left(S=h_2 \mid \mathbf{X}\right)} \frac{\delta^Y(\mathbf{X} ; h_1)}{\delta^D(\mathbf{X} ; h_1)}\left\{D-\mu_{D,0}\left(\mathbf{X} ; h_2\right)-\delta^D\left(\mathbf{X} ; h_2\right) Z\right\} \\
&+\frac{1}{\kappa}  \mathbbm{1}(S = t)\left\{\frac{\delta^Y(\mathbf{X} ; h_1)}{\delta^D(\mathbf{X} ; h_1)} \delta^D\left(\mathbf{X} ; h_2\right)-\psi\right\},
\end{aligned}}%
$$
where $\mu_{Y,z}\left(\mathbf{X}; s\right)=\mathbb{E}[Y\mid \mathbf{X}, S=s, Z=z]$, $\mu_{D,z}\left(\mathbf{X}; s\right)=\mathbb{E}[D\mid \mathbf{X}, S=s, Z=z]$ and $\kappa=f(S = t)$. The \textit{semiparametric efficiency bound} under  $\mathcal{M}$ is $\mathbb{E}[EIF_\psi^2]$.
\end{theorem}
 \bz{Although $\mathcal{M}$ is a non-parametric model, the treatment assignment probabilities $f(Z=z\mid \textbf{X}, S)$ are known by design in our setting and in particular do not typically depend on $\textbf{X}$. Nevertheless, it follows, e.g. from \citet{hahn1998role}, that knowledge of $f(Z=z \mid \textbf{X}, S)$ in this case should not change the bound.} In contrast, one may be able to leverage information on $f(S=s \mid \mathbf{X})$ to gain precision, although we do not pursue this since such knowledge is not generally available.


To construct an estimator of $\psi$ based on $EIF_\psi$, one must estimate $\delta^Y(\mathbf{X} ; h_1)$, $\delta^D(\mathbf{X} ; h_1)$ and $\delta^D(\mathbf{X};h_2)$, plus the additional nuisance functions $\mu_{Y,z}\left(\mathbf{X}; s\right)$, $\mu_{D,z}\left(\mathbf{X}; s\right)$ and $f(S=s\mid \mathbf{X})$. Although $f(Z\mid \textbf{X}, S=h_1)$ is known, one typically uses the \textit{estimated} version of it. One strategy would be to develop a multiply robust approach similar to that in \citet{wang2018bounded}, based on parametric working models for the nuisance functions. Here, we describe an alternative approach, which allows for off-the-shelf methods to learn these quantities. These could include classical non-parametric estimators (e.g. kernel smoothers, sieves) or potentially more flexible statistical learning  approaches (random forests, kernel ridge regression, Lasso, ensemble methods). After obtaining estimates $\hat{\delta}^Y(\mathbf{X} ; h_1)$, $\hat{\delta}^D(\mathbf{X} ; h_1)$, $\hat{\delta}^D(\mathbf{X};h_2)$, $\hat{f}(Z\mid \mathbf{X}, S=s)$, $\hat{f}(S=s\mid \mathbf{X})$, $\hat{\mu}_{Y,0}(\mathbf{X};h_1)$ and  $\hat{\mu}_{D,0}(\mathbf{X};h_2)$, one can then estimate $\psi$ as
 {\small 
\begin{align*}
 &\widehat{ITT}_{\text{EIF}} \\
= &\frac{1}{|\mathcal{D}_{t}|} \sum_{i = 1}^{|\mathcal{D}|}  \frac{(2 Z_i-1)  \mathbbm{1}\left(S_i=h_1\right)}{\hat{f}\left(Z_i \mid \mathbf{X}_i, S_i=h_1\right)} \frac{\hat{f}(S_i=t \mid \mathbf{X}_i)}{\hat{f}\left(S_i=h_1 \mid \mathbf{X}_i\right)} \frac{\hat{\delta}^D\left(\mathbf{X}_i; h_2\right)}{\hat{\delta}^D\left(\mathbf{X}_i; h_1\right)}\left[Y_i-\hat{\mu}_{Y,0}\left(\mathbf{X}_i; h_1\right)-\left\{D_i-\hat{\mu}_{D,0}\left(\mathbf{X}_i ; h_1\right)\right\}  \frac{\hat{\delta}^Y(\mathbf{X}_i ; h_1)}{\hat{\delta}^D(\mathbf{X}_i ; h_1)}\right]\\
  &+\frac{1}{|\mathcal{D}_{t}|} \sum_{i = 1}^{|\mathcal{D}|} \frac{(2 Z_i-1)  \mathbbm{1}\left(S_i=h_2\right)}{\hat{f}\left(Z_i \mid \mathbf{X}_i, S_i=h_2\right)} \frac{\hat{f}(S_i=t \mid \mathbf{X}_i)}{\hat{f}\left(S_i=h_2 \mid \mathbf{X}_i\right)} \frac{\hat{\delta}^Y(\mathbf{X}_i ; h_1)}{\hat{\delta}^D(\mathbf{X}_i ; h_1)}\left\{D_i-\hat{\mu}_{D,0}\left(\mathbf{X}_i ; h_2\right)-\hat{\delta}^D\left(\mathbf{X}_i; h_2\right) Z_i\right\}\\
  &+\frac{1}{|\mathcal{D}_{t}|} \sum_{i = 1}^{|\mathcal{D}|}\mathbbm{1}\{S_i = t\}\frac{\hat{\delta}^Y(\mathbf{X}_i ; h_1)}{\hat{\delta}^D(\mathbf{X}_i ; h_1)}\hat{\delta}^D\left(\mathbf{X}_i;h_2\right).
\end{align*}}%
Under regularity conditions, if each of the nuisance estimators converges to the truth with mean squared error rate shrinking faster than $n^{-1/4}$ and certain Donkser conditions on the nuisance functions hold \citep{van2000asymptotic}, then $\widehat{ITT}_{\text{EIF}}$ is $n^{1/2}$-consistent and asymptotically normal. Furthermore, supposing that $\mathbb E[D|\mathbf{X},S=h_1,Z]$ is consistently estimated, the estimator is asymptotically unbiased (although not necessarily $n^{1/2}$-consistent) so long as one of the following restrictions hold: (1) $\mathbb E[Y|\mathbf{X},S=h_1,Z]$ and $E[D|\mathbf{X},S=h_2,Z]$ are consistently estimated; (2) $\mathbb E[Y|\mathbf{X},S=h_1,Z]$ and $f(S|\mathbf{X})$ are consistently estimated; and (3) $\mathbb E[D|\mathbf{X},S=h_2,Z]$ and $f(S|\mathbf{X})$ are consistently estimated. See Section 4.5 of \citet{wang2018bounded} for a discussion about the robustness properties. Additional robustness may be attained by using doubly robust estimators for $\delta^Y\left(\mathbf{X};h_1\right)$, $\delta^D\left(\mathbf{X};h_1\right)$ and $\delta^D\left(\mathbf{X};h_2\right)$ and/or adopting the parametrizations in \citet{wang2018bounded}. An estimator of the asymptotic variance can be obtained using a sandwich estimator. As discussed in \citet{chernozhukov2017double}, if very flexible learning methods are used, sample-splitting (estimating the nuisance functions on a training split, and $\psi$ on a test split) or cross-fitting are recommended to alleviate the Donsker conditions. 

\vspace{-0.3cm}
\subsection{Estimation and inference in the \emph{post hoc} analysis}
\label{subsec: estimation in the post hoc stage}
\bz{The previous results straightforwardly extend to the setting where one wishes to evaluate the ITT effect of the AC versus placebo with data available from the target AC trial. In that case,  
 conditional compliance $\mathbb{E}[D(Z=1)|\mathbf{X},S=t]$ can be identified under randomization as $\mu_{D,1}(\mathbf{X};t):=\mathbb{E}[D|\mathbf{X},S=t,Z=1]$. However, $\mathbb{E}[D(Z=0)|\mathbf{X},S=t]$ cannot be identified as straightforwardly, because there is no placebo arm in the AC trial. We will proceed here, as in our case study, by treating $\mathbb{E}[D(Z=0)|\mathbf{X},S=t]$ as a sensitivity parameter $\mu^*_{D,0}(\mathbf{X};t)$, such that $\delta^{D*}(\mathbf{X};t):=\mu_{D,1}(\mathbf{X};t)-\mu^*_{D,0}(\mathbf{X};t)$ and $\hat{\delta}^{D*}(\mathbf{X};t):=\hat{\mu}_{D,1}(\mathbf{X};t)-\mu^*_{D,0}(\mathbf{X};t)$. In that case, the identification functional is now
$$\small
\psi=\mathbb{E}\left[\frac{\delta^Y(\mathbf{X} ; h_1)}{\delta^D(\mathbf{X} ; h_1)}\delta^{D*}(\mathbf{X};t) \mid S = t\right].
$$
Results on estimation follow closely along the lines described in the previous subsection. Indeed, the regression-based estimators equal
\begin{equation}
\begin{split}\small
    \widehat{ITT}_{\text{full, reg}} 
    &= \frac{1}{|\mathcal{D}_{t}|} \sum_{i = 1}^{|\mathcal{D}|} \mathbbm{1}\{S_i = t\}\times \frac{\hat{\delta}^Y(\mathbf{X}_i;h_1)}{\hat{\delta}^D(\mathbf{X}_i;h_1)}\hat{\delta}^{D*}(\mathbf{X}_i;t).
\end{split}
\end{equation}
whereas the estimators based on the efficient influence function simplify to
 {\small 
\begin{align*}
 &\widehat{ITT}_{\text{EIF}} \\
= &\frac{1}{|\mathcal{D}_{t}|} \sum_{i = 1}^{|\mathcal{D}|}  \frac{(2 Z_i-1)  \mathbbm{1}\left(S_i=h_1\right)}{\hat{f}\left(Z_i \mid \mathbf{X}_i, S_i=h_1\right)} \frac{\hat{f}(S_i=t \mid \mathbf{X}_i)}{\hat{f}\left(S_i=h_1 \mid \mathbf{X}_i\right)} \frac{\hat{\delta}^{D*}(\mathbf{X}_i;t)}{\hat{\delta}^D\left(\mathbf{X}_i; h_1\right)}\left[Y_i-\hat{\mu}_{Y,0}\left(\mathbf{X}_i; h_1\right)-\left\{D_i-\hat{\mu}_{D,0}\left(\mathbf{X}_i ; h_1\right)\right\}  \frac{\hat{\delta}^Y(\mathbf{X}_i ; h_1)}{\hat{\delta}^D(\mathbf{X}_i ; h_1)}\right]\\
  &+\frac{1}{|\mathcal{D}_{t}|} \sum_{i = 1}^{|\mathcal{D}|} \frac{\mathbbm{1}(Z_i=1)  \mathbbm{1}\left(S_i=t\right)}{\hat{f}\left(Z_i \mid \mathbf{X}_i, S_i=t\right)}  \frac{\hat{\delta}^Y(\mathbf{X}_i ; h_1)}{\hat{\delta}^D(\mathbf{X}_i ; h_1)}\left\{D_i-\hat{\mu}_{D,1}\left(\mathbf{X}_i ; t\right)\right\}+\frac{1}{|\mathcal{D}_{t}|} \sum_{i = 1}^{|\mathcal{D}|}\mathbbm{1}\{S_i = t\}\frac{\hat{\delta}^Y(\mathbf{X}_i ; h_1)}{\hat{\delta}^D(\mathbf{X}_i ; h_1)}\hat{\delta}^{D*}(\mathbf{X}_i;t).
\end{align*}}%
One can then estimate the ITT effect of the experimental therapy versus placebo in the target AC trial population by adding one of the estimates described above to the estimated ITT comparison of the experimental therapy versus active control. Although the above developments treat $\mu^*_{D,0}(\mathbf{X}_i;t)$ as fixed, in practice, one may wish to very it based on a plausible range of values.}

\vspace{-0.3cm}
\subsection{Extensions}
\bz{Point identification of $CATE(\mathbf{X}; h_1)$ requires imposing Assumption \ref{ass: no interaction} or other homogeneity-type assumptions on the historical trial $S = h_1$ \citep[Section 5.2]{swanson2018partial}. Alternatively, one may proceed by constructing partial identification intervals $[L(\mathbf{X}), U(\mathbf{X})]$ such that $CATE(\mathbf{X}; h_1) \in [L(\mathbf{X}), U(\mathbf{X})]$ almost surely. A partial identification interval bounds the range of possible values of the $CATE(\mathbf{X}; h_1)$ that are consistent with the observed data. Unlike a confidence interval, a partial identification interval would not shrink to a point even when the sample size goes to infinity, as the true parameter may take a range of values and cannot be point identified. Depending on the assumptions one is willing to make about the treatment assignment and treatment received, partial identification bounds with difference width can be formulated \citep{swanson2018partial}. We review some estimation strategies for partial identification bounds in Web Appendix B for completeness. We also construct partial identification bounds that are motivated by our case study in Section \ref{sec: case study}.} Web Appendix C also discusses some variants of the regression-based estimator and a sensitivity analysis assessing Assumption \ref{ass: no interaction}.


\vspace{-0.3cm}
\section{Simulation study}
\label{sec: simulation}
\vspace{-0.2cm}
\subsection{Goal and structure}
\label{subsec: simulation structure}
We consider data generating processes that have all three three sources of heterogeneity: treatment effect heterogeneity, within- and across-trial compliance heterogeneity, and target population heterogeneity. We generate two historical datasets $\mathcal{D}_{h_1}$ and $\mathcal{D}_{h_2}$, and a hypothetical placebo-controlled trial dataset $\mathcal{D}_{\text{target}}$ according to the following data generating process:

\begin{description}
\item[Sample sizes:] $N_1 = N_2 = N = 1000$, $2000$, and $5000$.

\item[\bz{Observed covariates and overlap:}] \bz{We consider the following two data generating processes for $\mathbf{X}$, one mimicking the case study (\textsf{Scenario X1}) and the other following a standard multivariate normal distribution (\textsf{Scenario X2}).

\textsf{Scenario X1:} We sample with replacement HPTN 084 participants' observed covariates to form $\mathcal{D}_{\text{target}}$. We then sample Partners PrEP participants' observed covariates to form $\mathcal{D}_{h_1}$ and $\mathcal{D}_{h_2}$ and control the amount of overlap between these two historical datasets and $\mathcal{D}_{target}$ using the following biased sampling strategy. For each study participant in Partners PrEP, we estimate a ``probability of trial participation," defined as the probability of selection into the HPTN 084 study over the Partners PrEP study based on a participant's baseline characteristics \citep{cole2010generalizing, Stuart:2011aa}. This ``probability of trial participation" is a version of \citeauthor{rosenbaum1983central}'s \citeyearpar{rosenbaum1983central} propensity score and captures the covariate balance between the target and historical datasets. By over- and under-sampling participants in Partners PrEP with large estimated ``probability of participation," we then control the amount of overlap between datasets. Specifically, the historical dataset $\mathcal{D}_{h_j}$ was formed by sampling $N_{j, \text{high}}$ and $N_{j, \text{low}}$ participants with high (above $0.5$) and low (below $0.5$) probability of participation, $j = 1, 2$. We consider three overlap levels: (i) Poor overlap: $N_{1, \text{high}}$ = 0.1$N_{1}$, $N_{1, \text{low}}$ = 0.9$N_{1}$, $N_{2, \text{high}}$ = 0.15$N_{2}$, $N_{2, \text{low}}$ = 0.85$N_{2}$; (ii) Limited overlap: $N_{1, \text{high}}$ = 0.19$N_{1}$, $N_{1, \text{low}}$ = 0.81$N_{1}$, $N_{2, \text{high}}$ = 0.19$N_{2}$, $N_{2, \text{low}}$ = 0.81$N_{2}$; (iii) Sufficient overlap: $N_{1, \text{high}}$ = 0.4$N_{1}$, $N_{1, \text{low}}$ = 0.6$N_{1}$, $N_{2, \text{high}}$ = 0.5$N_{2}$, $N_{2, \text{low}}$ = 0.5$N_{2}$. To illustrate, Figure S1 in Web Appendix D plots the overlap of the probability of participation between $\mathcal{D}_{\text{target}}$ and $\mathcal{D}_{h_1}$ in overlap, poor, and sufficient overlap scenarios when $N_1 = N = 2000$.} 

\textsf{Scenario X2:} We generate a 10-dimensional $\mathbf{X} \sim \text{Multivariate Normal}\left(\boldsymbol\mu, 0.5\cdot\mathbf {Id}\right)$, where $\mathbf {Id}$ is an identity matrix, $\boldsymbol \mu = (c, c, c, 0, \dots, 0)^{\text{T}}$ in $\mathcal{D}_{h_2}$, $\boldsymbol\mu = (1.2c, 1.2c, 1.2c, 0, \dots, 0)^{\text{T}}$ in $\mathcal{D}_{h_1}$, and $\boldsymbol\mu =$ $(0.8c, 0.8c, 0.8c, 0, \dots, 0)^{\text{T}}$ in $\mathcal{D}_{\text{target}}$, and $c \in \{0, 0.25, 0.50\}$. Parameter $c$ controls the amount of overlap in this scenario.

\item[Treatment assignment:] $Z$ is Bernoulli(0.5) in $\mathcal{D}_{h_1}$, $\mathcal{D}_{h_2}$ and $\mathcal{D}_{\text{target}}$.
\item[Treatment received:] $D$ is Bernoulli with $P(D(Z) = 1 \mid \mathbf{X}) = \text{expit}\{Z(2.5 -0.1X_1 + 0.3X_2 - 0.4X_5) + (1-Z)(-0.3X_1 - 0.4X_5 - 1.5)\}$ in $\mathcal{D}_{h_2}$ and $\mathcal{D}_{\text{target}}$, and $\text{expit}\{Z(2 + 0.2X_1 - 0.2X_5 ) + (1-Z)(-0.2X_5-1)\}$ in $\mathcal{D}_{h_1}$, \bz{where $\text{expit}(x) = \exp(x)/(1+\exp(x))$ is the inverse of the logit function.}
\end{description}
\noindent According to the above data generating process, the covariate distribution $\mathbf{X}$ is distinct among $\mathcal{D}_{h_1}$, $\mathcal{D}_{h_2}$, and $\mathcal{D}_{\text{target}}$ (that is, target population heterogeneity exists) in \textsf{Scenario X1} and in \textsf{Scenario X2} when $c \neq 0$. The effect of $Z$ on $D$ in $\mathcal{D}_{h_1}$ is different from that in $\mathcal{D}_{h_2}$ and $\mathcal{D}_{\text{target}}$ (that is, across-trial compliance heterogeneity exists). Moreover, $(X_1, X_2)$ modify the effect of $Z$ on $D$ in $\mathcal{D}_{h_2}$ and $\mathcal{D}_{\text{target}}$ (that is, within-trial compliance heterogeneity exists) so that the marginal compliance rate is different between $\mathcal{D}_{h_2}$ and $\mathcal{D}_{\text{target}}$. 

\begin{description}
    \item[Outcome:] \bz{We consider two sets of data generating processes in $\mathcal{D}_{h_1}$ and $\mathcal{D}_{h_2}$: a linear data generating process (\textsf{Scenario Y1}):
    \[\small
 P(Y(Z) = 1 \mid \mathbf{X}) =
 \begin{cases}
 \text{expit}\{Z(2.6 - 0.6X_1 - 0.8X_2 + 0.4X_3) + \\
 \ \ \ \ (1 - Z)(1.6 - 0.7X_1 - 0.7X_2 + 0.4X_3 - 0.2X_5)\}~\text{in}~\mathcal{D}_{h_1},\\
 \text{expit}\{1.4 - X_1 - 0.6X_2 + 0.4X_3 -0.6X_5 + 3.5Z\}~\text{in}~\mathcal{D}_{h_2},
 \end{cases}
 \]
and a nonlinear data generating process (\textsf{Scenario Y2}):
   \[\small
 P(Y(Z) = 1 \mid \mathbf{X}) =
 \begin{cases}
 \text{expit}\{Z(2.6 - 0.6X_1 - 0.8X_2 + 0.4\sqrt{X_3}) + \\
 \ \ \ \ (1 - Z)(1.6 - 0.7X_1 - 0.7X_2^3 + 0.4X_3 - 0.2X_5)\}~\text{in}~\mathcal{D}_{h_1},\\
 \text{expit}\{1.4 - X_1 - 0.6\sqrt{X_2} + 0.4X_3 -0.6X_5 + 3.5Z\}~\text{in}~\mathcal{D}_{h_2}.
 \end{cases}
 \]}
\end{description}
In the hypothetical trial dataset $\mathcal{D}_{\text{target}}$, we generated the potential outcome $P(Y(Z = 0) = 1 \mid \mathbf{X}) = 0$ and hence $P(Y(Z = 1) = 1 \mid \mathbf{X})$ equals the conditional intention-to-treat effect which is a product of the conditional average treatment effect in $\mathcal{D}_{h_1}$ and the conditional compliance in $\mathcal{D}_{h_2}$. The data-generating process also ensures that $CATE(\mathbf{X}; S = h_1)$ is bounded between $-1$ and $1$. 

We considered $6$ estimators of the intention-to-treat effect in the hypothetical placebo-controlled trial including (i) a difference-in-means estimator $\widehat{ITT}_{\text{hypo}}$ based on the unobservable outcome data in $\mathcal{D}_{\text{target}}$, (ii) and (iii) two covariate-adjusted estimators that (incorrectly) assume the conditional constancy assumption between $\mathcal{D}_{\text{target}}$ and $\mathcal{D}_{h_1}$ ($\widehat{ITT}_{\text{const}, 1}$) and between $\mathcal{D}_{\text{target}}$ and $\mathcal{D}_{h_2}$ ($\widehat{ITT}_{\text{const}, 2}$), (iv) a historical-data-driven, regression-based estimator $\widehat{ITT}_{\text{reg, par}}$, (v) a historical-data-driven, EIF-based estimator $\widehat{ITT}_{\text{EIF, par}}$ with all nuisance parameters estimated via parametric regression models, and (vi) a historical-data-driven, EIF-based estimator $\widehat{ITT}_{\text{EIF, gam}}$ with all nuisance parameters estimated via generalized additive models \citep{hastie2017generalized}. In each setting, we repeat the simulation $1000$ times. 

\subsection{Results}
\label{subsec: simulation results}
\bz{Figure S2 in Web Appendix D exhibits and compares the sampling distributions of $6$ estimators under consideration when the sample sizes are $N_1 = N_2 = N = 2000$, observed covariates are generated according to \textsf{Scenario X1}, and the outcomes are generated according to \textsf{Scenario Y1}. The ground truth intention-to-treat effects are superimposed using red dashed lines. The three historical-data-driven estimators $\widehat{ITT}_{\text{reg, par}}$,  $\widehat{ITT}_{\text{EIF, par}}$, and $\widehat{ITT}_{\text{EIF, gam}}$ all closely resemble the ground truth ITTs, though they have larger variances compared to that of the unobtainable, gold-standard estimator $\widehat{ITT}_{\text{hypo}}$. As the overlap between historical and target datasets improves, the variance of each historical-data-driven estimator starts to shrink and the sampling distribution becomes more concentrated around the ground truth. Table \ref{tbl: simulation results X1} summarizes the percentage of bias and coverage of $95\%$ confidence intervals for different sample sizes and overlap levels. We encountered simulated datasets where the estimator $\widehat{ITT}_{\text{EIF, gam}}$ became unstable due to the small weights in the denominator, especially when the covariate overlap is poor. In these cases, we applied a hard thresholding and let the estimator be $\phi(\widehat{ITT}_{\text{EIF, gam}})$ where function $\phi(x) = 1,~\forall x \geq 1$, $\phi(x) = -1,~\forall x \leq -1$, and $\phi(x) = x$ otherwise. The percentage of bias of $\widehat{ITT}_{\text{EIF, gam}}$ reported in Table \ref{tbl: simulation results X1} is based on the truncated version. Similar to the impression delivered by Figure S2, in all cases considered in this simulation study, three historical-data-driven estimators had small to negligible biases. On the other hand, two estimators based on the incorrect conditional constancy assumption ($\widehat{ITT}_{\text{const}, 1}$ and $\widehat{ITT}_{\text{const}, 2}$) were heavily biased and their confidence intervals' coverage was nowhere close to the nominal level. The bias that persists after adjusting for the observed covariates difference is often observed in empirical studies and referred to as "residual confounding" by \citet{zhang2014sensitivity}. Our simulation exhibits concrete settings where such residual confounding could emerge. The $95\%$ confidence intervals for all but $\widehat{ITT}_{\text{EIF, gam}}$ were based on nonparametric bootstrap. We found that the bootstrapped $95\%$ CIs of $\widehat{ITT}_{\text{reg, par}}$ and $\widehat{ITT}_{\text{EIF, par}}$ approximately attained their nominal level when sample sizes are as large as $2000$ in each dataset. The bootstrapped CIs of $\widehat{ITT}_{\text{EIF, gam}}$ were found to be highly conservative; on the other hand, the $95\%$ CIs obtained based on asymptotic normality and estimated asymptotic variances tended to undercover when the overlap was poor and the sample size was small, but began to achieve nominal coverage level when the overlap was sufficient and sample size was as large as $5000$. In the Web Appendix D, we report simulation results when observed covariates were generated according to \textsf{Scenario X2} and outcomes were generated according to \textsf{Scenario Y1} and \textsf{Scenario Y2}. In the additional nonlinear data generating process \textsf{Scenario Y2}, $\widehat{ITT}_{\text{EIF, gam}}$ continued to have negligible bias and good coverage, while $\widehat{ITT}_{\text{reg, par}}$ became biased once the parametric models became misspecified.}



\begin{table}[ht]
\centering
\resizebox{\textwidth}{!}{
\begin{tabular}{ccccccccccccc}
  \hline \\
    &\multicolumn{2}{c}{$\widehat{ITT}_{\text{hypo}}$}
    &\multicolumn{2}{c}{$\widehat{ITT}_{\text{const, 1}}$}
    &\multicolumn{2}{c}{$\widehat{ITT}_{\text{const, 2}}$} &\multicolumn{2}{c}{$\widehat{ITT}_{\text{reg, par}}$}
    &\multicolumn{2}{c}{$\widehat{ITT}_{\text{EIF, par}}$} 
    &\multicolumn{2}{c}{$\widehat{ITT}_{\text{EIF, gam}}$}\\
    \multirow{3}{*}{\begin{tabular}{c} Sample \\ size \end{tabular}} 
    &\multirow{3}{*}{\begin{tabular}{c}\%\\ Bias\end{tabular}}
    &\multirow{3}{*}{\begin{tabular}{c}\small$95\%$ CI \\ \small Coverage\end{tabular}}
    &\multirow{3}{*}{\begin{tabular}{c}\%\\ Bias\end{tabular}} 
    &\multirow{3}{*}{\begin{tabular}{c}\small$95\%$ CI \\ \small Coverage\end{tabular}}
    &\multirow{3}{*}{\begin{tabular}{c}\%\\ Bias\end{tabular}}
    &\multirow{3}{*}{\begin{tabular}{c}\small$95\%$ CI \\ \small Coverage\end{tabular}}
    &\multirow{3}{*}{\begin{tabular}{c}\%\\ Bias\end{tabular}}
    &\multirow{3}{*}{\begin{tabular}{c}\small$95\%$ CI \\ \small Coverage\end{tabular}}
    &\multirow{3}{*}{\begin{tabular}{c}\%\\ Bias\end{tabular}}
    &\multirow{3}{*}{\begin{tabular}{c}\small$95\%$ CI \\ \small Coverage\end{tabular}}
    &\multirow{3}{*}{\begin{tabular}{c}\%\\ Bias\end{tabular}}
    &\multirow{3}{*}{\begin{tabular}{c}\small$95\%$ CI \\ \small Coverage\end{tabular}}\\ \\ \\
  \hline 
  \multicolumn{13}{c}{Poor Overlap}\\ 
1000 
& 0.0  & 96.0\% & -16.5 & 93.2\% & 73.2  & 27.0\% & 2.5	& 96.0\% & 2.5 & 96.2\% & 2.7 & 84.0\% \\

2000 
& -0.1 & 97.4\% & -17.8  & 87.8\% & 72.1  &	5.0\% &	0.8  &	95.6\% & -0.7 &	92.8\% & -0.5 & 85.8\% \\ 

5000 
& -0.1 &	96.4\% & -17.0  & 81.0\%	& 70.4  & 0.0\% & 2.2 &	96.0\% & 2.3 &	94.9\%	& 2.3 &88.1\% \\ 

\multicolumn{13}{c}{Limited Overlap}\\
1000 
& -0.6 &	96.2\%	&-16.1	&91.2\%&	72.4&	19.2\%&	3.2 & 93.4\%	& 1.6	&94.4\%& 2.8	 & 89.6\% \\ 

2000 
& 0.4 &97.2\%	&-17.5 &85.6\% &	70.6 &3.8\%	& 1.3 &	95.6\%	&0.8 &94.6\%&	0.8	&88.6\% \\ 

5000 
& 0.3	& 95.3\%	&-17.0	&73.2\%	&72.4 &	0.0\%	& 2.0 &	94.7\%	& 2.3 &	95.1\%	& 2.3 &90.4\% \\

\multicolumn{13}{c}{Sufficient Overlap}\\
1000 
& 0.1 &	96.2\%&	-15.9 &	88.2\%&	72.2 &	5.6\%&	3.9& 94.6\%	&3.4 & 94.6\%&	4.2	&	92.8\% \\ 

2000 
& -0.7	&96.4\%	&-16.6	&83.0\%	&73.2&	0.0\%&	2.7	&94.0\%	&2.7 &93.8\%& 3.0	&90.8\% \\ 

5000 
& 0.1	&95.8\%	&-17.1	&62.2\%	&71.9 &0.0\%	&1.8	&94.7\%&	2.0	&95.2\%&	2.0 & 93.6\% \\

\hline
\end{tabular}}
\caption{\small Simulation results of $6$ estimators corresponding to {\small\textsf{Scenario X1}} and  {\small\textsf{Scenario Y1}}. The percentage of bias and coverage of $95\%$ confidence intervals are reported. Confidence intervals of $\widehat{ITT}_{\text{EIF, gam}}$ were estimated based on asymptotic normality and the efficient influence function. Confidence intervals of $\widehat{ITT}_{\text{hypo}}$ were based on two-sample tests. Confidence intervals of the other estimators were obtained via bootstrap.}
\label{tbl: simulation results X1}
\end{table}

\vspace{-0.3cm}
\section{Case study: Efficacy of daily TDF/FTC in HIV-1 prevention}\label{sec: case study}
\vspace{-0.2cm}
\subsection{Historical placebo-controlled trials of daily oral TDF/FTC}
\bz{Our goal is to estimate the $ITT$ effect of daily oral TDF/FTC versus placebo and then the $ITT$ effect of CAB-LA in the HPTN 084 trial population.} We consider an integrated analysis of the patient-level data from HPTN 084 and a historical placebo-controlled trial of daily oral TDF/FTC using our proposed framework and methods. There are 3 large-scale, multicenter, randomized trials that evaluated daily oral TDF/FTC: Partners PrEP \citep{baeten2012antiretroviral}, FEM-PrEP \citep{van2012preexposure}, and VOICE \citep{marrazzo2015tenofovir}. The FEM-PrRP and VOICE were conducted in the heterosexual women population in multiple African countries, while the Partners PrEP study enrolled HIV-uninfected heterosexual men and women who had a partner living with HIV (i.e., HIV-1–serodiscordant heterosexual couples) from Kenya and Uganda. These three historical studies recorded quite different annualized HIV incidence in the daily oral TDF/FTC arm. The Partners PrEP study reported an incidence of $0.95$ per $100$ person-years in the TDF/FTC arm among heterosexual women \citep[Figure 3]{baeten2012antiretroviral}. On the other hand, both the FEM-PrEP study \citep[Table 2]{van2012preexposure} and the VOICE study \citep[Table 3]{marrazzo2015tenofovir} reported an incidence of $4.7$ per $100$ person-years in the TDF/FTC arm. The annualized HIV incidence was reported to be $1.86$ per $100$ person-years in the TDF/FTC arm of the HPTN 084 study. In this integrated analysis, we chose the Partners PrEP study as the historical placebo-controlled trial because the gap between the annualized HIV incidence rate in the TDF/FTC arm between the Partners PrEP study and the HPTN 084 study, while still substantial, was considerably smaller compared to the other two studies. 

\vspace{-0.3cm}
\subsection{\bz{Overlap between the HPTN 084 and Partners PrEP trial populations; target population}}
\bz{We first examine the overlap between trial population of the HPTN 084 study and that of the Partners PrEP study. Because the HPTN 084 study enrolled only female participants, we focused on the female participants in the Partners PrEP study. Moreover, as the Partners PrEP study enrolled heterosexual women who had a partner living with HIV, the study would not reveal the ITT effect or the conditional average treatment effect of daily oral TDF/FTC for heterosexual women whose partners did not live with HIV, if a partner's HIV status is an important modifier of the compliance pattern or the treatment effect. Therefore, instead of making inference for the entire HPTN 084 population, we only focused on about one third of the HPTN 084 participants whose partners either living with HIV or having an unknown HIV status as our target population.

The second and third columns in Table \ref{Tableone_caseStudy} summarize the baseline characteristics of study participants in the target population of HPTN 084 and female participants in the Partners PrEP study. Compared to those in Partners PrEP, participants in the HPTN 084 target population were younger (mean age $26.3$ versus $33.5$) and received more education ($46.3\%$ versus $6.3\%$ completing the secondary school). HPTN 084 participants also had higher unemployment rate ($75.8\%$ versus $31.8\%$), higher positivity rates of baseline diagnoses of gonorrhea ($6.0\%$ versus $1.2\%$), chlamydia ($16.3\%$ versus $1.1\%$) and trichomonas ($7.8\%$ versus $6.8\%$), and lower positivity rate of syphilis ($2.6\%$ versus $5.8\%$). To help better summarize and visualize the covariate overlap between two population, Figure S6 in the Web Appendix E exhibits the distributions of the estimated ``probability of participation" in the target HPTN 084 population and among female participants in the Partners PrEP study \citep{cole2010generalizing,Stuart:2011aa}. The plot suggests that there is overlap between two populations across the spectrum of the ``probability of participation," although the overlap is limited so covariate adjustment is warranted.

\begin{table}[H]
\centering
\caption{\small Baseline characteristics of all participants in HPTN 084, HPTN 084 participants whose partners lived with HIV, and female participants in the Partners PrEP study. Mean (SD) are reported for continuous variables. Counts (\%) are reported for categorical variables.}
\label{Tableone_caseStudy}
\resizebox{0.7\textwidth}{!}{
\begin{tabular}[t]{lccc}
\toprule
  & \thead{\textbf{HPTN 084} \\ All \\participants} & \thead{\textbf{HPTN 084} \\ Participants whose \\ partner lived with HIV \\ or had an unknown status} & \thead{\textbf{Partners PrEP} \\ Female participants \\ whose partners \\lived with HIV} \\
\midrule
& (N=3224) & (N=1139) & (N=1184)\\
\addlinespace[0.3em]
\multicolumn{4}{l}{\textbf{Study arm}}\\
\hspace{0.5cm}CAB-LA & 1614 (50.1\%) & 559 (49.1\%) & 0 (0\%) \\
\hspace{0.5cm}TDF/FTC & 1610 (49.9\%) & 580 (50.9\%) & 565 (47.7\%)\\
\hspace{0.5cm}Placebo & 0 (0\%) & 0 (0\%) & 619 (52.3\%) \\
\addlinespace[0.3em]
\textbf{Age} & 26.0 (5.78) & 26.3 (6.03) & 33.5 (7.55)  \\
\addlinespace[0.3em]
\multicolumn{4}{l}{\textbf{Gonorrhea}}\\
\hspace{0.5cm}Neg & 2977 (92.3\%) & 1059 (93.0\%) & 1068 (90.2\%)\\
\hspace{0.5cm}Pos& 210 (6.5\%) & 68 (6.0\%) & 14 (1.2\%)  \\
\hspace{0.5cm}Missing & 37 (1.1\%) & 12 (1.1\%) & 102 (8.6\%) \\
\addlinespace[0.3em]
\multicolumn{4}{l}{\textbf{Chlamydia}}\\
\hspace{0.5cm}Neg &2583 (80.1\%) & 941 (82.6\%) & 1068 (90.2\%)\\
\hspace{0.5cm}Pos & 604 (18.7\%) & 186 (16.3\%) & 13 (1.1\%)  \\
\hspace{0.5cm}Missing & 37 (1.1\%) & 12 (1.1\%) & 103 (8.7\%) \\
\addlinespace[0.3em]
\multicolumn{4}{l}{\textbf{Trichomonas}}\\
\hspace{0.5cm}Neg & 2859 (88.7\%) & 1021 (89.6\%) & 1057 (89.3\%)\\
\hspace{0.5cm}Pos & 270 (8.4\%) & 89 (7.8\%) & 80 (6.8\%) \\
\hspace{0.5cm}Missing & 95 (2.9\%) & 29 (2.5\%) & 47 (4.0\%) \\
\addlinespace[0.3em]
\multicolumn{4}{l}{\textbf{Syphilis}}\\
\hspace{0.5cm}Neg& 3116 (96.7\%) & 1107 (97.2\%) & 1101 (93.0\%)\\
\hspace{0.5cm}Pos& 103 (3.2\%) & 30 (2.6\%) & 69 (5.8\%)  \\
\hspace{0.5cm}Missing & 5 (0.2\%) & 2 (0.2\%) & 14 (1.2\%) \\
\addlinespace[0.3em]
\multicolumn{4}{l}{\textbf{Employment}}\\
\hspace{0.5cm}Employed  & 878 (27.2\%) & 276 (24.2\%) & 807 (68.2\%) \\
\hspace{0.5cm}Not employed & 2346 (72.8\%) & 863 (75.8\%) & 377 (31.8\%)\\
\addlinespace[0.3em]
\multicolumn{4}{l}{\textbf{Education}}\\
\hspace{0.5cm}Complete secondary school  & 1528 (47.4\%) & 527 (46.3\%) & 75 (6.3\%)\\
\hspace{0.5cm}Not complete secondary school& 1346 (41.7\%) & 506 (44.4\%) & 271
(22.9\%) \\
\hspace{0.5cm}Not complete primary school & 350 (10.9\%) & 106 (9.3\%) & 838
(70.8\%)\\
\bottomrule
\end{tabular}}
\end{table}

The first column of Table \ref{Tableone_caseStudy} further exhibits the covariate distribution of the entire HPTN 084 study. We found that participants in the target population were similar to the entire HPTN 084 trial population in age, education, unemployment rate and baseline sexually transmitted infections; nevertheless, because partners' HIV status could be an important risk factor, we still restricted our analysis to $n = 1,139$ participants whose partners lives with HIV or had an unknown status. }

In addition to baseline characteristics of trial participants, self-reported adherence to the daily pill-taking was also different in HPTN 084 and Partners PrEP. Consuming at least $80\%$ of prescribed pills was typically considered ``adhering to the drug" in the HIV prevention literature \citep{murnane2015estimating}. Adopting this definition, $80.5\%$ of daily TDF/FTC recipients adhered to the prescription in the Partners PrEP Study, and this number was $52.4\%$ in the HPTN 084 study. In the HPTN 084 study, measurements of plasma tenofovir concentrations from a prespecified random cohort of $405$ study participants were obtained; $812$ out of $1,939$ samples ($41.0\%$) had tenofovir concentrations consistent with daily use ($\geq 40$ ng/mL). 

\begin{figure}[ht]
    \centering
    \includegraphics[width=\textwidth]{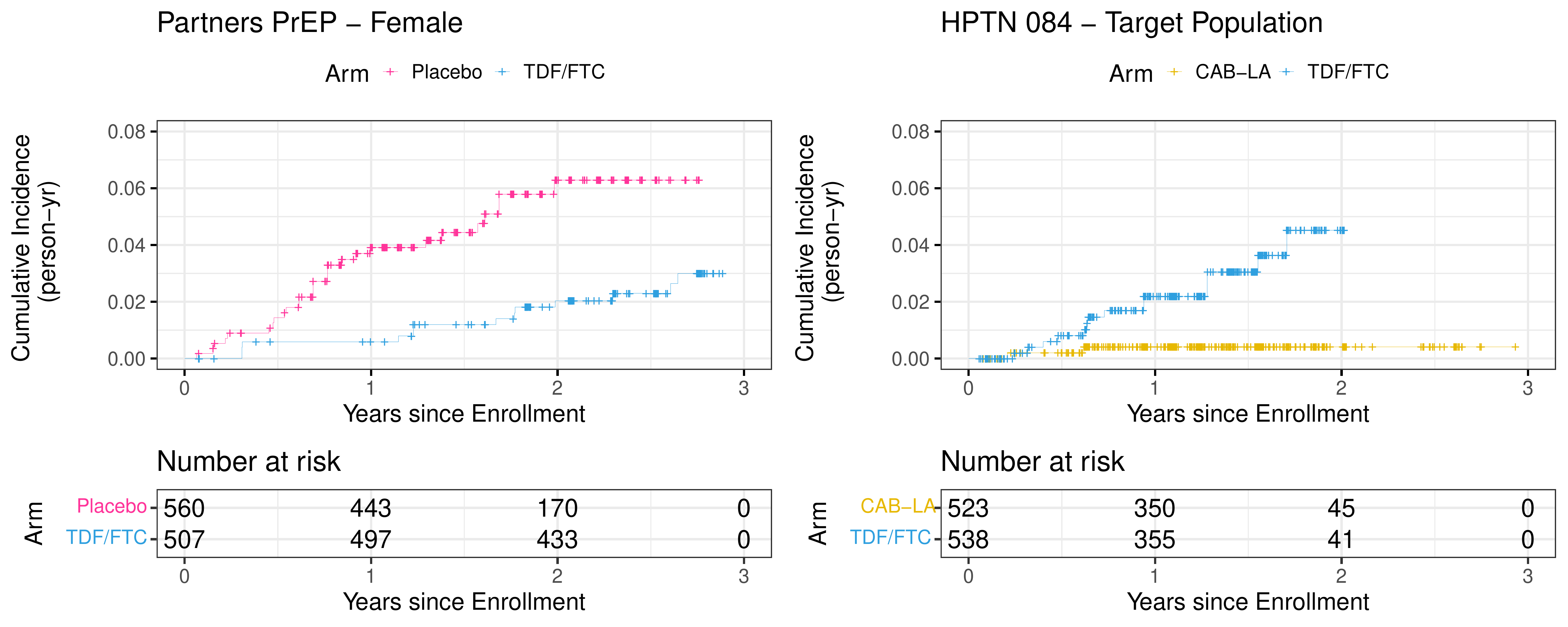}
    \caption{Left panel: Kaplan-Meier estimates of incident HIV acquisition in the Partners PrEP study. Right panel: Kaplan-Meier estimates of incident HIV acquisition among HPTN 084 participants whose partners either living with HIV or having an unknown HIV status (target population).}
    \label{fig: Incident HIV infection HPTN 084 and Partners PrEP}
\end{figure}

\bz{The right panel of Figure \ref{fig: Incident HIV infection HPTN 084 and Partners PrEP} plots the cumulative incidence curves in the CAB-LA and TDF/FTC arms in the target population, while the left panel plots the cumulative incidence curves in the TDF/FTC and placebo arms among heterosexual women in the Partners PrEP study. In view of the difference in patient composition and adherence pattern, both the constancy and conditional constancy assumptions are likely to be violated. Below, we seek to estimate the $ITT$ effect of daily oral TDF/FTC against placebo for the target population based on evidence from the Partners PrEP study using the framework developed in the article.} 

\vspace{-0.3cm}
\subsection{Estimating the ITT effect of daily oral TDF/FTC in the target population: two approaches}
\label{subsec: real data point identification}
We estimated the intention-to-treat effect of daily oral TDF/FTC against placebo in reducing HIV-1 incidence in the target population under both point and partial identification frameworks. First, we assume Assumption \ref{ass: no interaction} holds for the Partners PrEP study and the hypothetical placebo-controlled trial of daily oral TDF/FTC in the target population with the observed covariates including age, employment status, education, and four comorbidties including gonorrhea, chlamydia, trichomonas, and syphilis. Under this assumption and Assumption \ref{ass: mean generalizability}, we estimated the average treatment effect of daily oral TDF/FTC against placebo conditional on observed baseline covariates based on the Partners PrEP data. We then estimated the conditional compliance using the observed self-reported adherence data in the TDF/FTC arm in the HPTN 084 study and by treating the probability that a placebo recipient received daily oral TDF/FTC (i.e., $CC_{P}(\mathbf{X}; S = \text{HPTN 084})$) as a sensitivity parameter. In this way, if we assume no cross-over in the hypothetical placebo-controlled trial (i.e., $CC_{P}(\mathbf{X}; S = \text{HPTN 084}) = 0$), then the intention-to-treat effect of daily oral TDF/FTC against placebo in the target population was estimated to be $-3.9$ HIV infections per 100 person-years (95\% CI: $-9.7$ to $-0.7$). In a sensitivity analysis, we further allowed some minor degree of cross-over by setting $CC_{P}(\mathbf{X}; S = \text{HPTN 084}) = 5\%$ and $10\%$, and the ITT effect of TDF/FTC was estimated to be $-3.5$ per 100 person-years (95\% CI: $-8.7$ to $-0.6$) and $-3.1$ per 100 person-years (95\% CI: $-7.8$ to $-0.5$), respectively. \bz{We then conducted inference using the EIF-based estimator proposed in Section \ref{subsec: estimation in the post hoc stage}. The ITT effect of TDF/FTC was estimated to be $-3.1$ per 100 person-years (95\% CI: $-5.3$,  $-0.9$) assuming no cross-over. If we set $CC_{P}(\mathbf{X}; S = \text{HPTN 084}) = 5\%$ and $10\%$, the ITT effect of TDF/FTC was estimated to be $-2.5$ per 100 person-years (95\% CI: $-4.5$,  $-0.5$) and $-1.8$ per 100 person-years (95\% CI: $-3.6$,  $0.0$), respectively. In this integrated analysis, we found that the EIF-based estimators were more efficient compared to the regression-based estimators.}


\bz{Because the target population and the Partners PrEP population was not well-overlapped in baseline commodities including Gonorrhea and Chlamydia, we further considered an analysis restricted to participants testing negative for Gonorrhea, Chlamydia and Trichomonas at baseline. For this target population, the ITT effect of TDF/FTC against placebo was estimated to be $-3.3$ per 100 person-years (95\% CI: $-9.0$ to $-0.4$) assuming no cross-over. According to the EIF-based estimator, the ITT effect of TDF/FTC was estimated to be $-2.2$ per 100 person-years (95\% CI: $-4.5$ to $0.0$) assuming no cross-over.}

Finally, we considered relaxing Assumption \ref{ass: no interaction} in the Partners PrEP study and only partially identifying the conditional average treatment effect of daily TDF/FTC versus placebo. To this end, we considered the following strategy. Within each stratum defined by observed covariates $\mathbf{X}$, the conditional average treatment effect can be decomposed into a weighted sum of the treatment effect among the subgroup of compliers and that among the non-compliers, weighted by their relatively proportions in the stratum. It then suffices to determine the average treatment effect among the compliers, non-compliers, and their proportions, all within the strata of observed covariates. Conditional on the observed covariates, the complier average effect is identified by the ratio estimator \eqref{eqn: ratio estimator of CATE in S = h}. On the other hand, the treatment effect of daily oral TDF/FTC among non-compliers has the following natural bounds: the maximum treatment effect was to reduce all HIV incidence in the placebo arm of the Partners PrEP study and the minimum effect was $0$. In this way, we estimated bounds for the conditional average treatment effect and used these bounds to form the final ITT effect estimates against placebo in the target population. Assuming no cross-over, the interval estimates of the ITT effect were $[-3.6, -2.5]$ per 100 person-years (95\% CI: $-8.0$ to $-0.4$). 


\vspace{-0.3cm}
\subsection{Estimating the absolute efficacy of CAB-LA in the target population}
\label{subsec: real data CAB-LA}
Our analysis also immediately implies that the HIV incidence was $6.5$ (95\% CI: 3.1 to 12.4) per 100 person-years in the counterfactual placebo arm (primary analysis under the point identification assumptions and based on the regression-based estimator) in the target population. This estimate became $5.5$ (95\% CI: 1.5 to 11.0]) per 100 person-years if we further restrict the target population to those who tested negative for Gonorrhea, Chlamydia and Trichomonas at baseline. These estimates of placebo arm HIV incidence agreed reasonably well with those reported in the FEM-PrEP study ($5.0$ per 100 person-years) and the VOICE study ($4.6$ per 100 person-years). On the other hand, under a na\"ive adoption of the constancy assumption, one would conclude an HIV incidence of $3.7$ per 100 person-year in the target population, which appeared to largely underestimating the HIV incidence among young women in sub-Saharan Africa. Our result also implies an absolute efficacy of CAB-LA as large as $-6.1$ per 100 person-years (95\% CI: -11.9 to -2.6) in the target population and $-5.3$ per 100 person-years (95\% CI: -10.5 to -1.3) if the target population was further restricted to those who tested negative for Gonorrhea, Chlamydia and Trichomonas at baseline. Put together the estimates for HIV incidence in the placebo arm and the estimates of the efficacy of CAB-LA, \emph{we estimated that CAB-LA eliminated about 95\% of HIV acquisitions in the target population.}

\vspace{-0.3cm}
\section{Discussion}
\label{sec: discussion}
In this article, we systematically study the problem of generalizing the intention-to-treat effect of an active control versus placebo from historical placebo-controlled trials to an active-controlled trial. Our key insight is that generalization critically depends on the post-randomization event like adherence to the prescribed treatment in clinical trials. Our framework helps translate what FDA refers to as non-statistically-based uncertainties \citep[Page 20]{FDA2016} into concrete causal identification assumptions, highlights multiple sources of heterogeneity, including heterogeneity in participants composition, compliance, and treatment effect, and emphasizes the role of a post-randomization event when generalizing and transporting causal conclusions. 

Our work adds to existing HIV prevention literature on inferring intent-to-treat effect of an active-control based on a ``counterfactual placebo" incidence estimate, which may be constructed via leveraging data from a concurrent registrational cohort that receives access to available standard of care for HIV prevention \citep{PrEPVacc}, placebo arm data of historical trials in a similar population \citep{Donnell_CROI_2022}, HIV recency testing data collected at screening \citep{gao2021sample} and adherence-efficacy relationship \citep{glidden2020bayesian, glidden2021using}. 

The statistical problem of generalizing the ITT effect of an active control versus placebo from relevant historical trials has become more relevant as active-controlled trials have become increasingly prevalent. There are several ways to further this line of research. One important future direction is to generalize the framework to more complicated settings where post-randomization events like adherence to the intervention are time-varying, and the endpoint of interest is a time-to-event endpoint. Second, compliance or adherence to the intervention in an instrumental variable framework is a particular instance of a post-randomization event. It is also of interest to further extend the framework to a more generic, post-randomization event and allow a direct effect from the treatment to the endpoint of interest. \bz{Lastly, in many practical circumstances, researchers may not have the luxury to work with the patient-level data across multiple phase 3 clinical trials.
Study-level adherence from historical trials has been used in a meta-regression analysis to infer oral PrEP effectiveness \citep{hanscom2019adaptive}. Other meta-analysis-based approaches are also available; see, e.g., related discussion in Section \ref{subsec: intro literature review}. It is of interest to link the patient-level analysis proposed in this article to the meta-analysis-based framework and articulate what identification and modeling assumptions are needed to facilitate using only summary data from relevant historical trials.} 

\bz{Two statistical challenges are particularly relevant in generalizing efficacy estimates from historical data. First, researchers need to always pay close attention to the overlapping covariate space between the planned active-controlled trial and historical trials and, in our opinion, should always focus on the well-overlapped covariate space to avoid over-extrapolation with limited data. Traditional methods like multivariate matched sampling \citep{rubin1979using} can be generalized to the context of across-trials comparisons; see, e.g., \citet{zhang2023efficient}. Examining the scalar summary statistic like \citeauthor{Stuart:2011aa}'s \citeyearpar{Stuart:2011aa} ``probability of participation" is also useful. If the target trial enrolls a heterogeneous population of participants, then it is conceivable that multiple historical trials targeting different different constituent parts of the target population may be needed. Second, in some cases, it is conceivable that trials may not maintain a similar list of important covariates or may collect different versions of the same covariates. This is less of a concern if the studies were conducted via the same clinical trials network (e.g., the HIV Prevention Trials Network and the HIV Vaccine Trials Network) but could lead to many practical challenges and prevent researchers from pursuing covariate adjustment in other cases. Classical measurement error methods or methods that leverage proxy variables could be useful.} 

\bz{After decomposing the target estimand into a conditional compliance term ($CC(\mathbf{X})$) and a conditional average treatment effect term ($CATE(\mathbf{X})$), one relevant historical dataset is used to derive an estimate for this conditional average treatment effect term. In some scenarios, one may have access to multiple, patient-level historical datasets that may each inform an estimate of $CATE(\mathbf{X})$; one reasonable approach is to adopt a formal Bayesian framework as in \citet{zhou2019bayesian} or a random effect model to pool these $CATE(\mathbf{X})$ estimates.}

\vspace{-0.3cm}
\section*{Acknowledgement}
\label{sec: acknowledge}
We are grateful to the study participants, study staff and investigators on HPTN 084 and Partners PrEP who provided the data for this analysis.
We acknowledge the funders and sponsors of the trials. \bz{We are grateful to the HPTN Manuscript Review Committee for helpful feedback.} \bz{This work was supported by the U.S. National Institutes of Health grants R01AI177078 and UM1AI068617 (Fei Gao) and by the VIDD Faculty Initiative Award at the Fred Hutchinson Cancer Center (Fei Gao and Bo Zhang).} Oliver Dukes received support from the Research Foundation Flanders (1222522N).

\bibliographystyle{apalike}
\bibliography{paper-ref}

\begin{thebibliography}{}

\bibitem[Angrist et~al., 1996]{AIR1996}
Angrist, J.~D., Imbens, G.~W., and Rubin, D.~B. (1996).
\newblock Identification of causal effects using instrumental variables.
\newblock {\em Journal of the American Statistical Association}, 91(434):444--455.

\bibitem[Baeten et~al., 2012]{baeten2012antiretroviral}
Baeten, J.~M., Donnell, D., Ndase, P., Mugo, N.~R., Campbell, J.~D., Wangisi, J., Tappero, J.~W., Bukusi, E.~A., Cohen, C.~R., Katabira, E., et~al. (2012).
\newblock {Antiretroviral prophylaxis for HIV prevention in heterosexual men and women}.
\newblock {\em New England Journal of Medicine}, 367(5):399--410.

\bibitem[Bickel et~al., 1993]{bickel1993efficient}
Bickel, P.~J., Klaassen, C.~A., Bickel, P.~J., Ritov, Y., Klaassen, J., Wellner, J.~A., and Ritov, Y. (1993).
\newblock {\em Efficient and adaptive estimation for semiparametric models}, volume~4.
\newblock Springer.

\bibitem[Cheng and Huang, 2010]{cheng2010bootstrap}
Cheng, G. and Huang, J.~Z. (2010).
\newblock Bootstrap consistency for general semiparametric m-estimation.
\newblock {\em The Annals of Statistics}, 38(5):2884--2915.

\bibitem[Chernozhukov et~al., 2017]{chernozhukov2017double}
Chernozhukov, V., Chetverikov, D., Demirer, M., Duflo, E., Hansen, C., and Newey, W. (2017).
\newblock Double/debiased/neyman machine learning of treatment effects.
\newblock {\em American Economic Review}, 107(5):261--65.

\bibitem[Cohen and Baden, 2012]{cohen2012preexposure}
Cohen, M.~S. and Baden, L.~R. (2012).
\newblock Preexposure prophylaxis for {HIV}—where do we go from here?
\newblock {\em New England Journal of Medicine}, 367(5):459--461.

\bibitem[Cole and Stuart, 2010]{cole2010generalizing}
Cole, S.~R. and Stuart, E.~A. (2010).
\newblock Generalizing evidence from randomized clinical trials to target populations: the actg 320 trial.
\newblock {\em American journal of epidemiology}, 172(1):107--115.

\bibitem[Dahabreh et~al., 2022]{dahabreh2022generalizing}
Dahabreh, I.~J., Robertson, S.~E., and Hern{\'a}n, M.~A. (2022).
\newblock Generalizing and transporting inferences about the effects of treatment assignment subject to non-adherence.
\newblock {\em arXiv preprint arXiv:2211.04876}.

\bibitem[Dahabreh et~al., 2019]{Dahabreh:2019aa}
Dahabreh, I.~J., Robertson, S.~E., Tchetgen, E.~J., Stuart, E.~A., and Hern{\'a}n, M.~A. (2019).
\newblock Generalizing causal inferences from individuals in randomized trials to all trial-eligible individuals.
\newblock {\em Biometrics}, 75(2):685--694.

\bibitem[Degtiar and Rose, 2021]{degtiar2021review}
Degtiar, I. and Rose, S. (2021).
\newblock A review of generalizability and transportability.
\newblock {\em arXiv preprint arXiv:2102.11904}.

\bibitem[Delany-Moretlwe et~al., 2022]{delany2022cabotegravir}
Delany-Moretlwe, S., Hughes, J.~P., Bock, P., Ouma, S.~G., Hunidzarira, P., Kalonji, D., Kayange, N., Makhema, J., Mandima, P., Mathew, C., et~al. (2022).
\newblock Cabotegravir for the prevention of {HIV}-1 in women: results from {HPTN} 084, a phase 3, randomised clinical trial.
\newblock {\em The Lancet}, 399(10337):1779--1789.

\bibitem[Donnell et~al., 2022]{Donnell_CROI_2022}
Donnell, D., Gao, F., Hughes, J., and Hanscom, B. (2022).
\newblock Counterfactual estimation of {CAB-LA} efficacy against placebo using external trials.
\newblock volume~86, Virtual.

\bibitem[Ellenberg and Temple, 2000]{ellenberg2000placebo}
Ellenberg, S.~S. and Temple, R. (2000).
\newblock Placebo-controlled trials and active-control trials in the evaluation of new treatments. part 2: practical issues and specific cases.
\newblock {\em Annals of Internal Medicine}, 133(6):464--470.

\bibitem[Fauci, 2017]{fauci2017hiv}
Fauci, A.~S. (2017).
\newblock An hiv vaccine is essential for ending the hiv/aids pandemic.
\newblock {\em Jama}, 318(16):1535--1536.

\bibitem[Fleming et~al., 2011]{fleming2011some}
Fleming, T.~R., Odem-Davis, K., Rothmann, M.~D., and Li~Shen, Y. (2011).
\newblock Some essential considerations in the design and conduct of non-inferiority trials.
\newblock {\em Clinical Trials}, 8(4):432--439.

\bibitem[{Food and Drug Administration}, 2016]{FDA2016}
{Food and Drug Administration} (2016).
\newblock {Non-inferiority clinical trials to establish effectiveness: Guidance for industry}.

\bibitem[Gao et~al., 2021]{gao2021sample}
Gao, F., Glidden, D.~V., Hughes, J.~P., and Donnell, D.~J. (2021).
\newblock Sample size calculation for active-arm trial with counterfactual incidence based on recency assay.
\newblock {\em Statistical Communications in Infectious Diseases}, 13(1).

\bibitem[Glidden et~al., 2021]{glidden2021using}
Glidden, D.~V., Das, M., Dunn, D.~T., Ebrahimi, R., Zhao, Y., Stirrup, O.~T., Baeten, J.~M., and Anderson, P.~L. (2021).
\newblock Using the adherence-efficacy relationship of emtricitabine and tenofovir disoproxil fumarate to calculate background hiv incidence: a secondary analysis of a randomized, controlled trial.
\newblock {\em Journal of the International AIDS Society}, 24(5):e25744.

\bibitem[Glidden et~al., 2020]{glidden2020bayesian}
Glidden, D.~V., Stirrup, O.~T., and Dunn, D.~T. (2020).
\newblock A bayesian averted infection framework for prep trials with low numbers of hiv infections: application to the results of the discover trial.
\newblock {\em The Lancet HIV}, 7(11):e791--e796.

\bibitem[Grant et~al., 2014]{grant2014uptake}
Grant, R.~M., Anderson, P.~L., McMahan, V., Liu, A., Amico, K.~R., Mehrotra, M., Hosek, S., Mosquera, C., Casapia, M., Montoya, O., et~al. (2014).
\newblock Uptake of pre-exposure prophylaxis, sexual practices, and hiv incidence in men and transgender women who have sex with men: a cohort study.
\newblock {\em The Lancet infectious diseases}, 14(9):820--829.

\bibitem[Hahn, 1998]{hahn1998role}
Hahn, J. (1998).
\newblock On the role of the propensity score in efficient semiparametric estimation of average treatment effects.
\newblock {\em Econometrica}, pages 315--331.

\bibitem[Hanscom et~al., 2019]{hanscom2019adaptive}
Hanscom, B., Hughes, J.~P., Williamson, B.~D., and Donnell, D. (2019).
\newblock Adaptive non-inferiority margins under observable non-constancy.
\newblock {\em Statistical methods in medical research}, 28(10-11):3318--3332.

\bibitem[Hastie, 2017]{hastie2017generalized}
Hastie, T.~J. (2017).
\newblock Generalized additive models.
\newblock In {\em Statistical models in S}, pages 249--307. Routledge.

\bibitem[Hern{\'a}n et~al., 2017]{hernan2017per}
Hern{\'a}n, M.~A., Robins, J.~M., et~al. (2017).
\newblock Per-protocol analyses of pragmatic trials.
\newblock {\em New England Journal of Medicine}, 377(14):1391--1398.

\bibitem[James~Hung et~al., 2003]{james2003some}
James~Hung, H., Wang, S.-J., Tsong, Y., Lawrence, J., and O'Neil, R.~T. (2003).
\newblock Some fundamental issues with non-inferiority testing in active controlled trials.
\newblock {\em Statistics in Medicine}, 22(2):213--225.

\bibitem[Joffe and Greene, 2009]{joffe2009related}
Joffe, M.~M. and Greene, T. (2009).
\newblock Related causal frameworks for surrogate outcomes.
\newblock {\em Biometrics}, 65(2):530--538.

\bibitem[Marrazzo et~al., 2015]{marrazzo2015tenofovir}
Marrazzo, J.~M., Ramjee, G., Richardson, B.~A., Gomez, K., Mgodi, N., Nair, G., Palanee, T., Nakabiito, C., Van Der~Straten, A., Noguchi, L., et~al. (2015).
\newblock {Tenofovir-based preexposure prophylaxis for HIV infection among African women}.
\newblock {\em New England Journal of Medicine}, 372(6):509--518.

\bibitem[Miner et~al., 2021]{miner2021broadly}
Miner, M.~D., Corey, L., and Montefiori, D. (2021).
\newblock Broadly neutralizing monoclonal antibodies for hiv prevention.
\newblock {\em Journal of the International AIDS Society}, 24:e25829.

\bibitem[Murnane et~al., 2015]{murnane2015estimating}
Murnane, P.~M., Brown, E.~R., Donnell, D., Coley, R.~Y., Mugo, N., Mujugira, A., Celum, C., Baeten, J.~M., Team, P. P.~S., Mujugira, A., et~al. (2015).
\newblock Estimating efficacy in a randomized trial with product nonadherence: application of multiple methods to a trial of preexposure prophylaxis for {HIV} prevention.
\newblock {\em American Journal of Epidemiology}, 182(10):848--856.

\bibitem[Neilan et~al., 2022]{neilan2022cost}
Neilan, A.~M., Landovitz, R.~J., Le, M.~H., Grinsztejn, B., Freedberg, K.~A., McCauley, M., Wattananimitgul, N., Cohen, M.~S., Ciaranello, A.~L., Clement, M.~E., et~al. (2022).
\newblock Cost-effectiveness of long-acting injectable hiv preexposure prophylaxis in the united states: a cost-effectiveness analysis.
\newblock {\em Annals of internal medicine}, 175(4):479--489.

\bibitem[Neyman, 1923]{neyman1923application}
Neyman, J.~S. (1923).
\newblock {On the application of probability theory to agricultural experiments. Essay on principles. Section 9.}
\newblock {\em Annals of Agricultural Sciences}, 10:1--51.

\bibitem[Pearl, 2011]{pearl2011transportability}
Pearl, J. (2011).
\newblock Transportability across studies: A formal approach.

\bibitem[Rosenbaum and Rubin, 1983]{rosenbaum1983central}
Rosenbaum, P.~R. and Rubin, D.~B. (1983).
\newblock The central role of the propensity score in observational studies for causal effects.
\newblock {\em Biometrika}, 70(1):41--55.

\bibitem[Rothmann et~al., 2003]{rothmann2003design}
Rothmann, M., Li, N., Chen, G., Chi, G.~Y., Temple, R., and Tsou, H.-H. (2003).
\newblock Design and analysis of non-inferiority mortality trials in oncology.
\newblock {\em Statistics in Medicine}, 22(2):239--264.

\bibitem[Rubin, 1980]{rubin1980discussion}
Rubin, D. (1980).
\newblock {Discussion of ``Randomization analysis of experimental data in the Fisher randomization test" by D. Basu}.
\newblock {\em Journal of the American Statistical Association}, 75:591--593.

\bibitem[Rubin, 1974]{rubin1974estimating}
Rubin, D.~B. (1974).
\newblock Estimating causal effects of treatments in randomized and nonrandomized studies.
\newblock {\em Journal of Educational Psychology}, 66(5):688.

\bibitem[Rubin, 1979]{rubin1979using}
Rubin, D.~B. (1979).
\newblock Using multivariate matched sampling and regression adjustment to control bias in observational studies.
\newblock {\em Journal of the American Statistical Association}, 74(366a):318--328.

\bibitem[Rudolph and van~der Laan, 2017]{Kara2016}
Rudolph, K.~E. and van~der Laan, M.~J. (2017).
\newblock Robust estimation of encouragement design intervention effects transported across sites.
\newblock {\em Journal of the Royal Statistical Society: Series B (Statistical Methodology)}, 79(5):1509--1525.

\bibitem[Stefanski and Boos, 2002]{stefanski2002calculus}
Stefanski, L.~A. and Boos, D.~D. (2002).
\newblock The calculus of {M}-estimation.
\newblock {\em The American Statistician}, 56(1):29--38.

\bibitem[Stuart et~al., 2011]{Stuart:2011aa}
Stuart, E.~A., Cole, S.~R., Bradshaw, C.~P., and Leaf, P.~J. (2011).
\newblock The use of propensity scores to assess the generalizability of results from randomized trials.
\newblock {\em Journal of the Royal Statistical Society. Series A, (Statistics in Society)}, 174(2):369--386.

\bibitem[Swanson et~al., 2018]{swanson2018partial}
Swanson, S.~A., Hern{\'a}n, M.~A., Miller, M., Robins, J.~M., and Richardson, T.~S. (2018).
\newblock Partial identification of the average treatment effect using instrumental variables: review of methods for binary instruments, treatments, and outcomes.
\newblock {\em Journal of the American Statistical Association}, 113(522):933--947.

\bibitem[{US National Library of Medicine}, 2021]{PrEPVacc}
{US National Library of Medicine} (2021).
\newblock {A Combination Efficacy Study in Africa of Two DNA-MVA-Env Protein or DNA-Env Protein HIV-1 Vaccine Regimens With PrEP (PrEPVacc)}.

\bibitem[Van~Damme et~al., 2012]{van2012preexposure}
Van~Damme, L., Corneli, A., Ahmed, K., Agot, K., Lombaard, J., Kapiga, S., Malahleha, M., Owino, F., Manongi, R., Onyango, J., et~al. (2012).
\newblock {Preexposure prophylaxis for HIV infection among African women}.
\newblock {\em New England Journal of Medicine}, 367(5):411--422.

\bibitem[van~der Laan and Rose, 2011]{van2011targeted}
van~der Laan, M.~J. and Rose, S. (2011).
\newblock {\em Targeted learning: causal inference for observational and experimental data}, volume~10.
\newblock Springer.

\bibitem[Van~der Vaart, 2000]{van2000asymptotic}
Van~der Vaart, A.~W. (2000).
\newblock {\em Asymptotic statistics}, volume~3.
\newblock Cambridge university press.

\bibitem[Wang and Tchetgen~Tchetgen, 2018]{wang2018bounded}
Wang, L. and Tchetgen~Tchetgen, E. (2018).
\newblock Bounded, efficient and multiply robust estimation of average treatment effects using instrumental variables.
\newblock {\em Journal of the Royal Statistical Society: Series B (Statistical Methodology)}, 80(3):531--550.

\bibitem[{World Health Organization}, 2022]{world2022guidelines}
{World Health Organization} (2022).
\newblock {\em Guidelines on long-acting injectable cabotegravir for {HIV} prevention}.
\newblock World Health Organization.

\bibitem[Zhang, 2023]{zhang2023efficient}
Zhang, B. (2023).
\newblock Efficient algorithms for building representative matched pairs with enhanced generalizability.
\newblock {\em Biometrics (in press)}.

\bibitem[Zhang, 2009]{zhang2009covariate}
Zhang, Z. (2009).
\newblock Covariate-adjusted putative placebo analysis in active-controlled clinical trials.
\newblock {\em Statistics in Biopharmaceutical Research}, 1(3):279--290.

\bibitem[Zhang et~al., 2014]{zhang2014sensitivity}
Zhang, Z., Nie, L., Soon, G., and Zhang, B. (2014).
\newblock Sensitivity analysis in non-inferiority trials with residual inconstancy after covariate adjustment.
\newblock {\em Journal of the Royal Statistical Society: Series C (Applied Statistics)}, 63(4):515--538.

\bibitem[Zhou et~al., 2019]{zhou2019bayesian}
Zhou, J., Hodges, J.~S., Suri, M. F.~K., and Chu, H. (2019).
\newblock A bayesian hierarchical model estimating cace in meta-analysis of randomized clinical trials with noncompliance.
\newblock {\em Biometrics}, 75(3):978--987.

\bibitem[Zhou et~al., 2022]{zhou2022estimating}
Zhou, T., Zhou, J., Hodges, J.~S., Lin, L., Chen, Y., Cole, S.~R., and Chu, H. (2022).
\newblock Estimating the complier average causal effect in a meta-analysis of randomized clinical trials with binary outcomes accounting for noncompliance: A generalized linear latent and mixed model approach.
\newblock {\em American journal of epidemiology}, 191(1):220--229.

\end{thebibliography}

\end{document}